\documentclass{ar-1col}

\usepackage[comma]{natbib}
\usepackage{url}
\usepackage[english]{babel}
\usepackage{amsmath,amssymb, amsthm, enumerate}
\usepackage{graphicx,subcaption}
\usepackage{tikz}
\usetikzlibrary{patterns}
\jname{Xxxx. Xxx. Xxx. Xxx.}
\jvol{AA}
\jyear{YYYY}
\doi{10.1146/((please add article doi))}

\definecolor{green_new}{rgb}{0, .5, 0}
\definecolor{blue_new}{rgb}{0, 0.3, 1.0}
\definecolor{red_new}{rgb}{0.9, 0, 0.2}

\newtheorem{example}{Example}
\newtheorem{remark}{Remark}

\def\g{\boldsymbol}
\def\ci{\perp\!\!\!\perp}
\def\perpe{\perp_e}
\DeclareMathOperator{\argmin}{arg\,min}

\DeclareMathOperator{\ph}{ph}

\DeclareMathOperator{\MST}{mst}

\newcommand{\HR}{H\"usler--Reiss}

\newcommand{\einsfun}{\g 1} 
\newcommand{\LL}{\mathcal L}
\newcommand{\EE}{\mathcal E}
\newcommand{\p}{\mathbb{P}}
\newcommand{\e}{\mathbb{E}}
\newcommand{\pa}{\mathrm{pa}}

\newcommand{\eps}{\varepsilon}
\newcommand{\convd}{\stackrel{\rm d}{\to}}

\newcommand{\R}{\mathbb{R}}

\newcommand{\E}{\mathbb{E}}

\begin{document}
\markboth{S.~Engelke and J.~Ivanovs}{Sparse Structures for Multivariate Extremes}

\title{Sparse Structures for Multivariate Extremes}

\author{Sebastian Engelke$^1$ and Jevgenijs Ivanovs$^2$
\affil{$^1$Research Center for Statistics, University of Geneva, Boulevard du Pont d’Arve 40, 1205 Geneva, Switzerland; email: sebastian.engelke@unige.ch}
\affil{$^2$Department of Mathematics, Aarhus University, Ny Munkegade 118, 8000 Aarhus, Denmark; email: jevgenijs.ivanovs@math.au.dk}}

\begin{abstract}
  Extreme value statistics provides accurate estimates for the small occurrence probabilities of rare events. While theory and statistical tools for univariate extremes are well-developed, methods for high-dimensional and complex data sets are still scarce. Appropriate notions of sparsity and connections to other fields such as machine learning, graphical models and high-dimensional statistics have only recently been established. This article reviews the new domain of research concerned with the detection and modeling of sparse patterns in rare events. We first describe the different forms of extremal dependence that can arise between the largest observations of a multivariate random vector. We then discuss the current research topics including clustering, principal component analysis and graphical modeling for extremes. Identification of groups of variables which can be concomitantly extreme is also addressed. The methods are illustrated with an application to flood risk assessment.  
\end{abstract}

\maketitle

\tableofcontents

\section{Introduction}

Flooding, heat waves and high concentrations of pollutants in the air are examples of environmental risks that are driven by very few rare events. Such events can have devastating impact on human life and can cause huge physical damage. Recent financial crises have likewise shown how underestimating the tails of loss distributions can underestimate systemic economic risks. The accurate statistical assessment of the small probabilities of occurrence of such extreme scenarios is thus crucial in many different settings. Extreme value theory is a widely-used approach to quantify the risk of these rare events. It provides mathematically justified tools to extrapolate beyond the data range and to estimate return periods of events that have never yet been observed.

In complex systems, such as rivers or financial networks, the most catastrophic events are due to concatenations of several rare events. Inundation of a lower river basin is typically the result of cumulation of simultaneous river exceedances in the upper river basin \citep[e.g.,][]{kee2009, asadi2015extremes}. 
In climate science, extreme impacts such as fires or droughts are driven by joint extremes of several meteorological variables \citep[e.g.,][]{wes2011,zsc2017, eng2017a}.
Similarly, the systemic risk of a financial system highly depends on the connections among core institutions \citep[e.g.,][]{poo2004, zho2010, mcn2015}. In all these applications, the multivariate dependence between univariate rare events will determine the severity of risk for the whole system.
Multivariate extreme value statistics therefore concentrates on dependence modeling in complex multivariate or spatial systems \citep[see][]{dav2012b}. While research is very active in this area, most applications are still limited to fairly moderate dimensions due to a lack of clear notions of sparsity in this context.

This review describes the existing literature, recent advances and
future directions in the mathematical theory and statistical methodology for
modeling dependence and detecting sparse patterns for extremes in higher dimensions.

\subsection{Overview}
The definition of extreme values implies that only few observations in a data set contain an informative signal on the distributional tail. Research on multivariate extremes in the last decades has thus concentrated on parsimonious modeling in cases where domain knowledge is available. A major branch with many applications in meteorology is the analysis of spatial extreme events, where information on the geographical locations of measurement stations significantly simplifies extremal dependence modeling.
In many applications, however, such knowledge is insufficient
or even unavailable, as for instance in risk analysis of financial networks where connections between the institutions are unknown. Especially in higher dimensions and complex situations it therefore becomes essential to exploit underlying structures and to learn sparse patterns in a data driven way.

Dependence between extreme observations of a random vector $X = (X_1,\dots, X_d)$ can exhibit complicated structures \citep[see][]{led1997, col1999} and the notions of sparsity, conditional independence and dimension reduction are sometimes different from the non-extreme world. Much recent work in extreme value statistics has started to establish links to other fields such as graphical models, machine learning and causality, and to adapt classical methods for the detection of sparse structures in multivariate data.
The different approaches can be grouped into three broad areas of research.

\begin{itemize}
\item[(i)]
  The first class of approaches concentrates on methods from unsupervised learning such as clustering and principal component analysis, and adapts them to the context of extreme observations. These non-parametric dimension reduction techniques are mostly used for exploratory analysis and visualization of extremal dependence \citep[see][]{cha2015, coo2019, dre2019, jan2019}. 
\item[(ii)]
  The second notion of sparsity is inherent to rare event analysis. It relates to the study of concomitant extremes, that is, which sub-groups of variables in the multivariate random vector $(X_1,\dots, X_d)$ are likely to take large values simultaneously. Sparse models should only exhibit a small number of such groups
  and their detection is a challenging task. Statistically this is related to estimating the support of a measure on a $d$-dimensional space and several inferential methods have been proposed \citep[see][]{goi2017,chi2017, chi2019, mey2019, sim2018}. 
\item[(iii)]
  One classical way to define probabilistic sparsity is through conditional independence structures
  and graphical models, since they allow the decomposition of high-dimensional distributions into low-dimensional components. Graphical models in extreme value statistics have only recently been introduced and studied \citep[see][]{gis2018, eng2018, seg2019}. These developments open new fields of research at the interface of extremes, structure learning, high-dimensional inference and causality \citep[see][]{mha2019,gne2019, eng2020}.
\end{itemize}

This review begins with some background on the fundamental objects of multivariate extreme value theory in Section~\ref{sec:prel}. The classical statistical modeling strategies and their limitations are briefly described in Section~\ref{sec:classical}.
Sections~\ref{sec:adapt}, \ref{sec:faces} and \ref{sec:markov} discuss the three main research directions for sparsity detection outlined above. Each time, a clear definition of the sparsity notion used in the respective section is given.

Our review conveys the main ideas in sparse modeling of extremes, but the literature is vast and our references are necessarily selective. Further interesting topics that are beyond the scope of this article include the modeling of asymptotically independent extremes \citep[e.g.,][]{HeffernanTawn2004, wadsworth2012dependence, pap2017}, flexible models linking different dependence classes \citep[e.g.,][]{Wadsworthetal2017, HuserWadsworth2017, eng2018a} and connections to the theory of networks \citep[e.g.,][]{sam2016, wan2020}.

\subsection{Application to flood risk assessment in Switzerland}\label{river}

\begin{figure}
	\centering
	\includegraphics[width=1\textwidth]{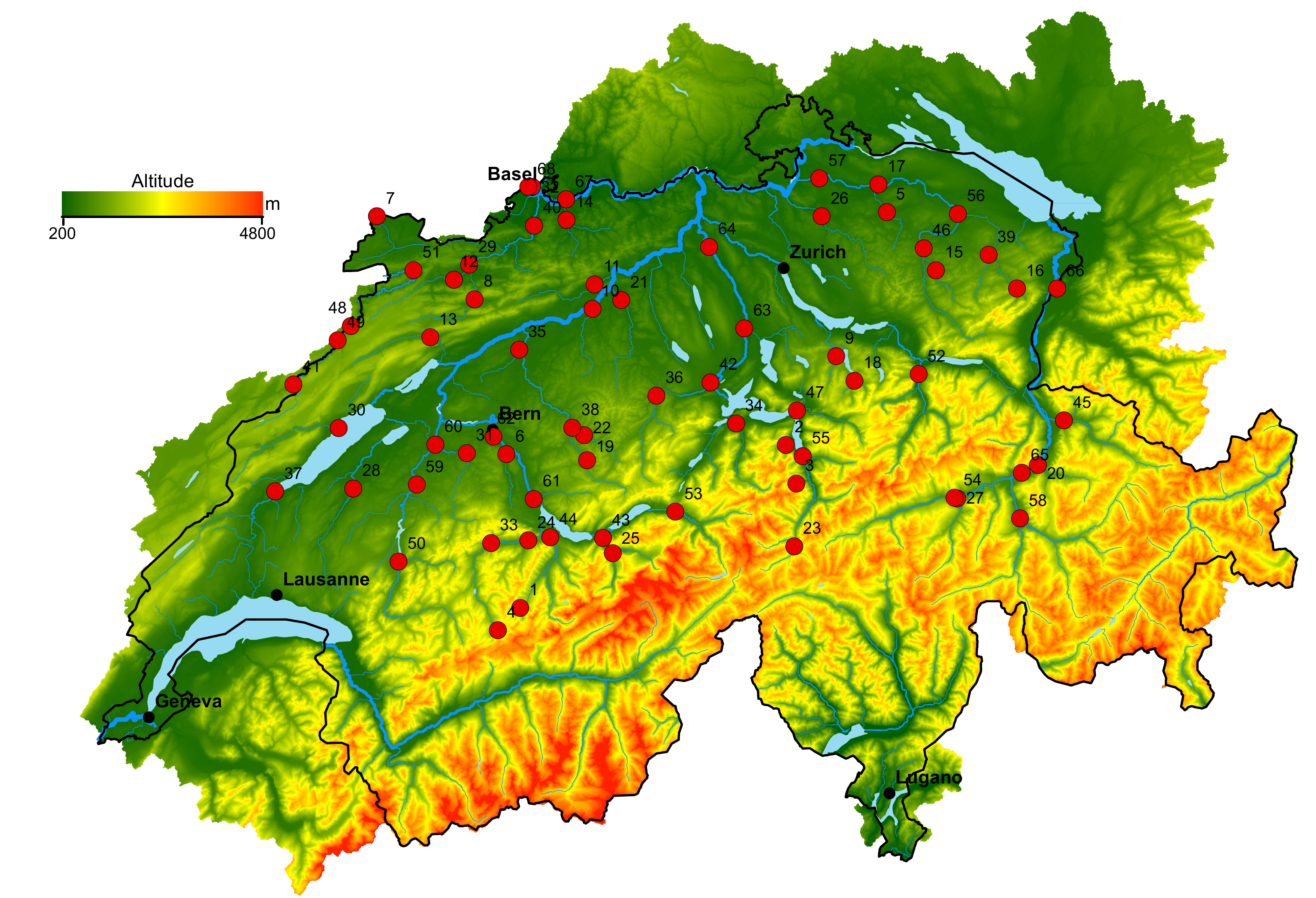}
	\caption{Topographic map of Switzerland showing sites of 68 gauging stations (red dots) mostly along the Rhine, the Aare and their tributaries.}
	\label{fig:basin}
\end{figure}

We illustrate the different methods of this review on sparse structures for extremes using river discharges
at $d=68$ locations in Switzerland, mostly in the Rhine and Aare catchments.
Figure~\ref{fig:basin} shows the basin with its topography and the gauging stations.
The data are monitored by the Swiss Federal Office of the
Environment and consist of daily average discharges. The length of the recorded time series at the 68 locations is between 30 and 120 years. For simplicity we only use the summer months June, July and August, and only data with records for all locations. This results in $22$ years of common summer discharges, that is, $n=2024$ daily observations.

Accurate quantification of the risk related to large peak river flows is crucial for
effective flood protection. A univariate extreme value analysis of the tails at each of the 68 stations has been done in \citet{asa2018}. Analyzing the extremal dependence structure of river discharges requires a wide range of statistical tools.
This includes the identification of groups of locations where floods may happen simultaneously, the statistical modeling of these concomitant extremes and the simulation of multivariate rare events for worst case analyses.

River networks are highly complex systems and the dependence between extremes at different locations can not be sufficiently explained by spatial Euclidean distances as is common in geostatistical applications for precipitation, for instance. In addition, the largest discharges may be dampened by big lakes or affected by hydroelectric installations.
For this data set on river flows it is thus highly relevant to understand the extremal dependence and we expect sparse patterns and a non-trivial underlying probabilistic structure.

\section{Preliminaries}\label{sec:prel}
\subsection{Recap of univariate theory}\label{sec:uni}
Univariate extreme value theory is a well-established topic and many statistical tools exist for analysis of the tail behavior of a random variable $X$. In the most basic setting, given independent observations $X^{(1)}, \dots, X^{(n)}$ of $X$, we are interested in estimating the survival function $1-F(t)=\p(X>t)$ for large~$t$. Here `large' is understood as being close to the maximal possible value $t_\infty=\sup\{t:F(t)<1\}$, known as the upper endpoint of~$F$.

There are two main modeling strategies based on different limiting probability models:
the block maxima method and the peaks-over-threshold approach. For the former, we assume that the sequence of normalized maxima converges in distribution to some non-degenerate limit,
\begin{align}\label{eq:GEV}
\frac{\max(X_1,\dots , X_n)-b_n}{a_n}\convd Z, \qquad n\to\infty, 
\end{align}
for some $a_n>0,b_n\in\R$. The distribution of $Z$ belongs to the class of generalized extreme value distributions~\citep{fis1928}, which is parameterized by its shape, location and scale parameters. 
Such $Z$ is also called max-stable because the maximum of independent copies of $Z$ can be normalized to get back the distribution of~$Z$.
The convergence in Equation~\ref{eq:GEV} is equivalent to the convergence of scaled exceedances to a non-degenerate limit~\citep{bal74},
\begin{align}\label{eq:GPD}
\left.\frac{X-t}{c_t} \, \right|\, \{X>t\}\convd Y,\qquad t\to t_\infty,
\end{align}
for some $c_t>0$, which underlies the peaks-over-threshold approach. The limit $Y$ has a generalized Pareto distribution ~\citep{Pickands1975}, parameterized by a shape and a location parameter.
Importantly, $Z$ and~$Y$ are closely related and they share the same shape parameter.

For a detailed overview of univariate extreme value theory and a further references we refer to the textbooks \citet{emb1997, coles2001introduction, ber2004, deh2006a} and \citet{res2008}, and to the review articles~\citet{kat2002} and \citet{dav2015}.

\subsection{Extremal dependence coefficients}\label{sec:coef}
Consider a $d$-dimensional random vector $X=(X_j: j \in V)$, where here and in the sequel $V = \{1,\dots, d\}$ denotes the index set. Our interest is in the probability that some (or all) components of $X$ are large. This probability is strongly influenced by the dependence between the extreme observations of the $d$ single variables.  Extremal dependence  may take many different forms.
For two components $X_i$ and $X_j$, a first broad split can be done through the
(upper) tail dependence coefficient, which is defined as 
\begin{align}\label{eq:chi}
 \chi_{ij} = \lim_{q\to 1} \chi_{ij}(q) = \lim_{q\to 1}\p(F_i(X_i)>q,F_j(X_j)>q) / (1-q) \in[0,1],
\end{align}
whenever the limit exists and where $F_i$ is the distribution function of $X_i$. It quantifies the conditional probability that both components are large given that one is large.
If the coefficient $\chi_{ij}>0$, the variables  $X_i$ and $X_j$ are said
to exhibit asymptotic dependence.
In the case $\chi_{ij} = 0$ we have asymptotic independence, and then one often assumes that
\begin{align}
  \p(F_i(X_i)>q,F_j(X_j)>q) = (1-q)^{1/\eta_{ij}}\ell(1-q), \qquad \eta_{ij} \in [0,1], \label{eq:eta}
\end{align}
where the measurable function $\ell:[0,1] \to \R_+$ is slowly varying at zero, that is, $\lim_{t\to 0}\ell(st)/\ell(t)=1$ for all $s>0$.
The coefficient $\eta_{ij}$ is called residual tail dependence coefficient, introduced by~\cite{led1997} and studied in~\cite{pen99,ram09,deHaa11} and \cite{Eas12}.
 It describes the rate of convergence of the joint exceedance probability to zero, and in the case of asymptotic dependence we have~$\eta_{ij} = 1$. For most bivariate distributions the coefficients $\chi_{ij}$ and $\eta_{ij}$ can be computed explicitly \citep[e.g.,][]{eng2018a}.

We may extend the definition of both tail dependence coefficients in Equations~\ref{eq:chi} and~\ref{eq:eta} to any non-empty subset $I\subset V=\{1,\ldots,d\}$ by considering joint exceedances of the components $X_i$, $i \in I$, and we denote them by $\chi_I$ and $\eta_I$.
The set of coefficients $\chi_I$ and $\eta_I$ for all non-empty $I\subsetneq V$ must satisfy the consistency constraint 
\begin{align}
&\sum_{J\supset I} (-1)^{|J\backslash I|}\chi_J\geq 0,
&\forall J\supset I\quad \eta_J\leq \eta_I.\label{eq:chi_eta_ineq} 
 \end{align} 
 Conversely, any such vectors $(\chi_I)$ and $(\eta_I)$ with elements in $[0,1]$ and $\chi_I>0$ implying $\eta_I=1$ can arise as tail dependence coefficients for some $d$-dimensional vector~$X$. 
Equation~\ref{eq:chi_eta_ineq} further implies the monotonicity $\chi_J\leq \chi_I$ for all $J\supset I$.
The above consistency result essentially follows from~\citet{deHaa11}; see also~\citet{schlather02} and \citet{strokorb15} for some further theory.

\begin{figure}
  \centering
    \begin{subfigure}[b]{0.49\textwidth}
  \includegraphics[width=\textwidth]{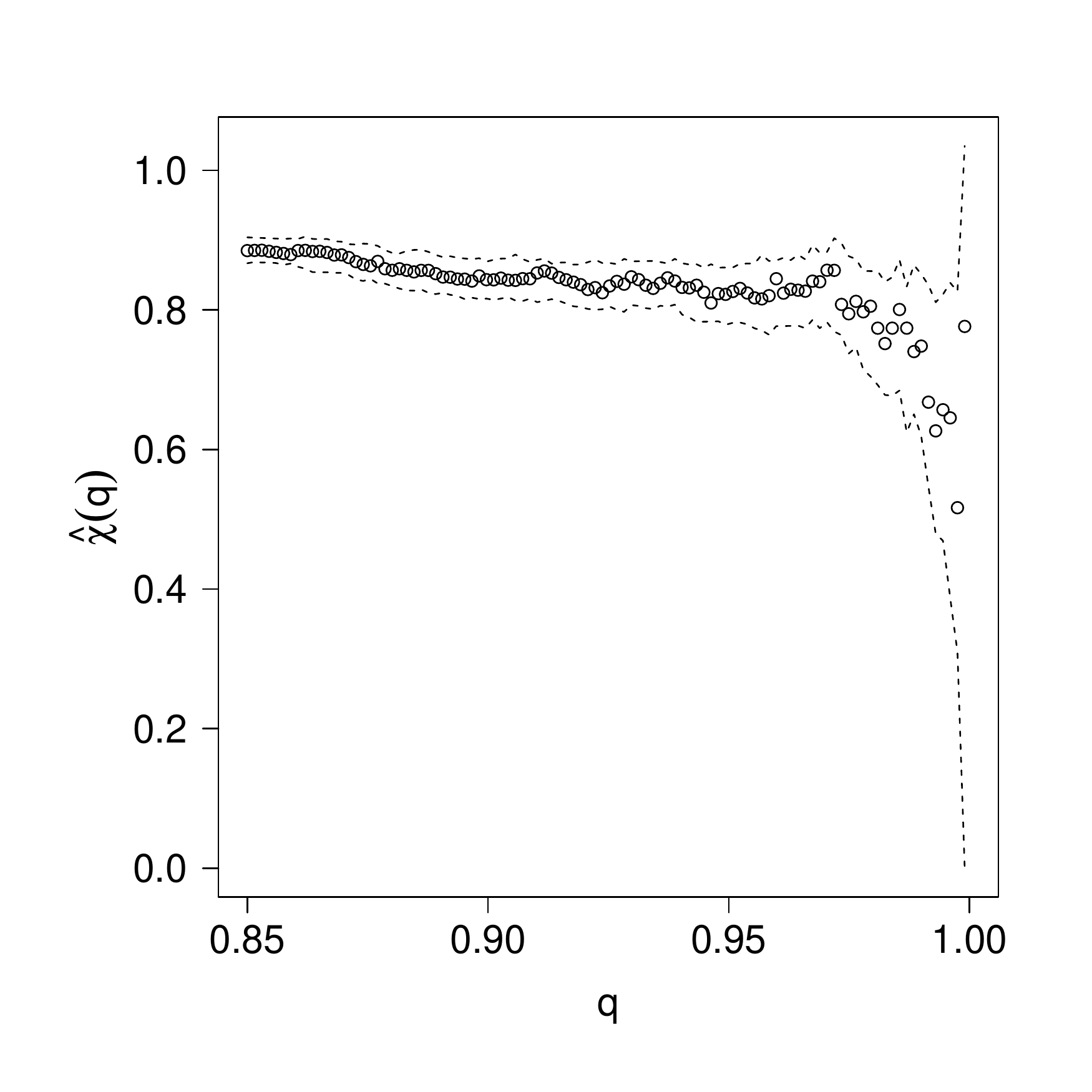}
  \end{subfigure}
      \begin{subfigure}[b]{0.49\textwidth}
  \includegraphics[width=\textwidth]{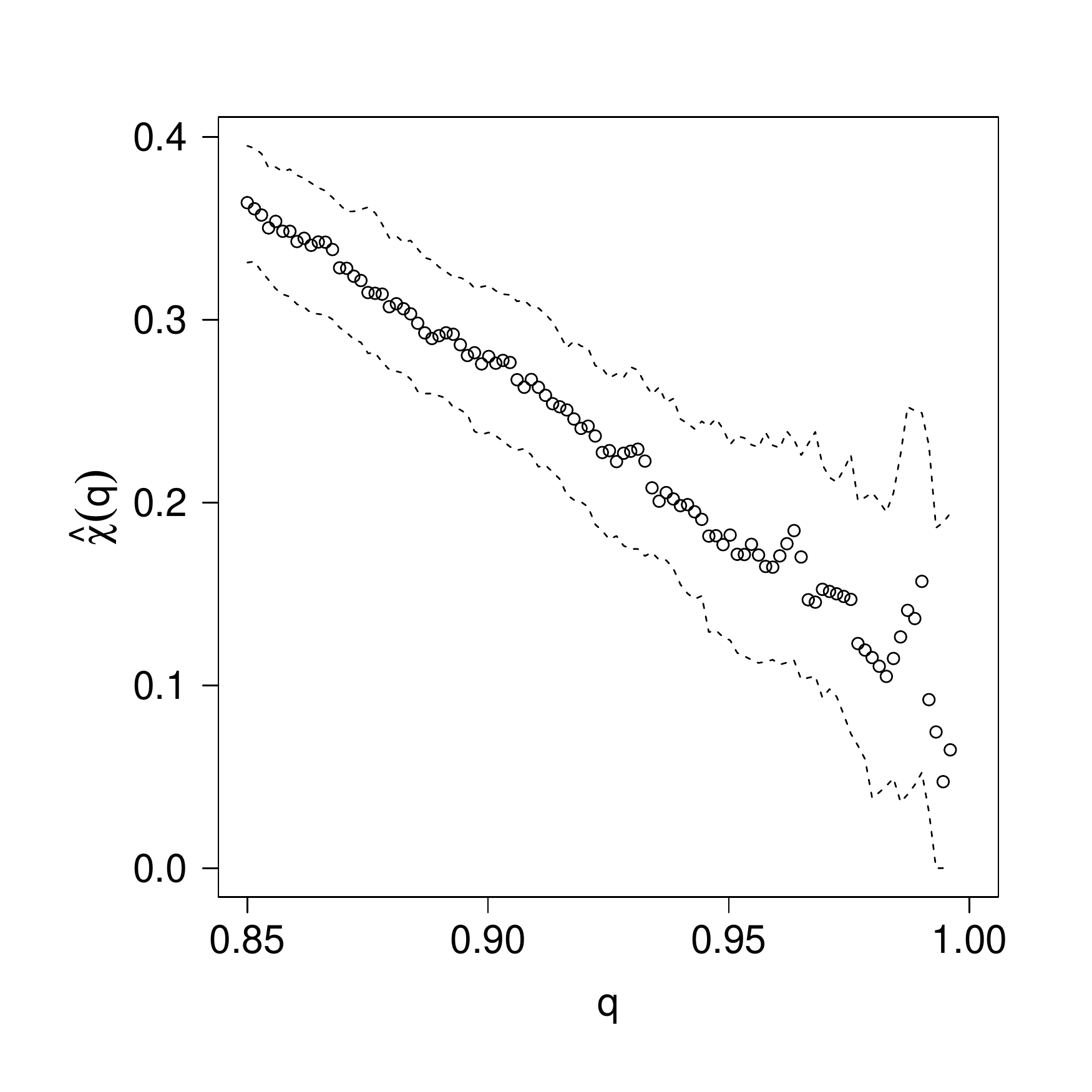}
  \end{subfigure}
  \caption{Empirical estimates $\widehat \chi_{ij}(q)$ of the tail dependence coefficient for a range of thresholds $q$ for $i=56$ and $j=57$ (left), and $i=1$ and $j=56$ (right). Dashed lines are $95\%$ bootstrap confidence intervals.}
  \label{fig:chi_plot}
\end{figure}

The coefficients presented here are summaries of the extremal dependence of the vector~$X$.
For a multivariate data set, a first exploratory analysis includes plots of empirical estimates
of the bivariate coefficients $\widehat \chi_{ij}(q)$ for a range of threshold levels $q$ close to one.
This helps to distinguish between the regimes of asymptotic dependence and independence and guides later modeling choices. For the Swiss river data from Section~\ref{river}, Figure~\ref{fig:chi_plot} shows
such plots for two pairs of stations. The curve in left-hand side plot corresponding to two close-by stations is stable around a positive level, indicating asymptotic dependence. The curve in the right-hand side plot corresponds to two stations far apart, and it tends to zero for $q\to 1$, which suggests asymptotic independence.

\subsection{Multivariate regular variation}\label{sec:mrv}

In multivariate extremes the problem
of analyzing the tail of the vector $X$ is usually divided into two steps, modeling of marginal tails and modeling of the extremal dependence. While the former step is described in Section~\ref{sec:uni}, for the latter we standardize the marginals to focus exclusively on extremal dependence.
The common choice is the standard Pareto distribution and we assume in the sequel that $\p(X_j\leq x)=1 - 1/x$ for $x\geq 1$ and $j\in V$.
For a continuous marginal distribution $F_j$ this amounts to a simple transformation of $X_j$ to $1/\{1-F_j(X_j)\}$. This procedure brings all the components to the same scale, so that `large' is now understood in the same way. In practice, this transformation can be done empirically (see Section~\ref{sec:stat}), or based on a parametric or semi-parametric estimate of $F_j$.

Similarly to the univariate setting, in multivariate extreme value theory there exist two
intimately linked methods to study the tail of the random vector $X$, namely the maxima approach and the peaks-over-threshold approach.
For the former, we consider component-wise maxima $M_n=\max(X^{(1)},\dots, X^{(n)})$ of independent copies of the random vector~$X$, that is, the $j$th component is $M_{nj} = \max_{i=1,\dots, n} X^{(i)}_{j}$. We assume 
that $M_n$ weakly converges as $n\to \infty$, when properly normalized, to some random vector~$Z$; see also Equation~\ref{eq:GEV} for the univariate case.
Since $X$ has standard Pareto margins, the normalization simplifies and we have $M_n/n\convd Z$,
and the marginals of $Z$ are standard Fr\'echet. The random vector $Z$ is called max-stable and its distribution can be represented as
\begin{equation}\label{eq:mu_z}
 \p(Z\leq z)=\exp\{-\Lambda_z\}=\exp\left\{-\Lambda\big(\mathcal E\setminus [0,z]\big)\right\},\qquad z\geq 0,
 \end{equation}
where the so-called exponent measure $\Lambda$ defined on the space $\mathcal E=[0,\infty)^d\setminus \{0\}$ satisfies $\Lambda(A)<\infty$ for all Borel sets $A\subset \mathcal E$ bounded away from the origin.
A standard argument shows that the above convergence of normalized maxima is equivalent to
\begin{equation}\label{eq:mu}
\lim_{t\to\infty}t\p(X/t\in A)=\Lambda(A),
\end{equation}
for all $\Lambda$-continuous Borel sets $A\subset \mathcal E$ bounded away from the origin.
The regularity property in Equation~\ref{eq:mu} is called multivariate regular variation.
Importantly, it suggests a simple way of extrapolating the probability law from, say, moderately large values into tail regions having few or no observations.

While under multivariate regular variation the componentwise maxima converge to the max-stable $Z$ defined in Equation~\ref{eq:mu_z}, the exceedances over a high threshold
converge to a multivariate Pareto distribution $Y$  \citep[see][]{roo2006}, 
\begin{align}
\label{mpd}
\p( Y \leq z ) &= \lim_{t\to \infty} \p\left(X / t \leq   z \mid \| X\|_\infty > t\right) =\frac{\Lambda_{ \min(z,1)} - \Lambda_z}{\Lambda_1}, \qquad z \in \LL,
\end{align}
where the support $\LL = \{x \in \mathcal E: \|x\|_\infty \geq  1\}$ is the positive orthant with the unit cube removed, and $\|z\|_\infty=\max_i |z_i|$ denotes the $\ell_\infty$-norm of $z \in \mathbb R^d$. This approximation follows from Equation~\ref{eq:mu} and it also implies that the law of $Y$ is proportional to $\Lambda$ restricted to $\mathcal L$.

Throughout the paper we assume that $X$ is multivariate regularly varying as defined in Equation~\ref{eq:mu}.

\subsection{Properties of the exponent measure}\label{sec:props}

The exponent measure $\Lambda$ contains all information on the extremal dependence of $X$.
Equation~\ref{eq:mu} immediately implies that it is homogeneous of order $-1$, that is, $\Lambda(c A) = c^{-1} \Lambda(A)$ for $c>0$.
It is often convenient to switch to polar coordinates, and thus we consider some norm $\|\cdot \|$; the usual choices are the $\ell_1$-norm $\|z\|_1=\sum_i|z_i|$ and $\ell_\infty$-norm. Define the positive simplex $\mathbb S^{d-1}_+=\{z\in \mathcal E:\|z\|=1\}$, so that each $z\in \mathcal E$ can be written as $z=\|z\|\theta$, where $\theta=z/\|z\|\in\mathbb S^{d-1}_+$ is the corresponding angle.
Homogeneity implies that $\Lambda$ decomposes into an angular part and an independent radial part
\begin{align}\label{spectral_decomp}
  \Lambda\{z\in \mathcal E:\|z\|\geq r,z/\|z\|\in \cdot\}=cr^{-1}\p(\Theta\in\cdot),\qquad \forall\, r>0,
\end{align}
where $c=\Lambda\{z\in \mathcal E:\|z\|>1\}$ is a fixed constant and $\Theta$ follows the so-called angular (or spectral) distribution $H$ on $\mathbb S^{d-1}_+$.
From Equation~\ref{eq:mu} we also have
\begin{equation}\label{eq:mrv}
\lim_{t\to \infty} \p\left(\frac{X}{\|X\|}\in\cdot \, \Big| \, \|X\|>t\right) =  \p(\Theta\in \cdot),
\end{equation}
leading to the interpretation of $\Theta$ as the limiting extremal angle for high threshold exceedances.
For a textbook treatment of multivariate regular variation we refer to~\cite{res2008}, and to~\citet{basrak2002characterization} and \citet{lindskog14} for further theory.

\begin{figure}
\centering
\includegraphics{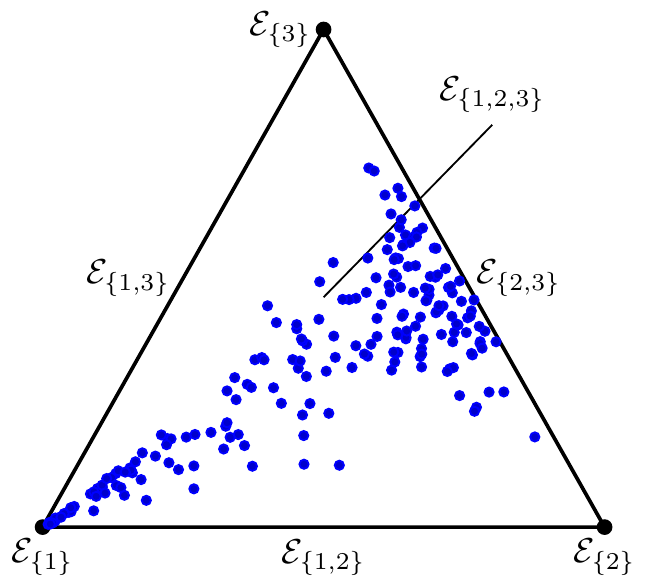}
\caption{The simplex $\mathbb S^2_+$ with empirical extremal angles corresponding to the sub-group $\{1,56,57\}$ of stations.}
\label{fig:simplex}
\end{figure}

Under multivariate regular variation all tail dependence coefficients in Equation~\ref{eq:chi} exist and $\chi_I=\Lambda(x_i>1, \forall i\in I)$.
Groups of components $(X_i:i\in I)$ that can be large simultaneously correspond to the non-empty subsets $I\subset V$ with $\chi_I>0$.
We can partition $\EE$ into $2^d-1$ disjoint sub-cones, the faces of all dimensions,
\begin{align}\label{E_I}
\EE_I &= \left\{x \in \EE: x_i > 0 \,\forall i\in I, x_j=0\, \forall j\notin I\right\},
\end{align}
and we note that $\Lambda(\EE_I)>0$
indicates that the components $(X_i:i\in I)$ can be extreme while the components
$(X_i:i\notin I)$ are much smaller.
Equivalently, this can be formulated in terms of mass of the multivariate Pareto distribution $Y$ 
or the angular measure $H$ on the corresponding faces. 
In particular, mass of $\Lambda$ in $\EE_V$, the interior of $\EE$, means that all components can be extreme at the same time. 
In principle, almost any collection of faces may have $\Lambda$-mass, resulting in order of $2^{2^d}$ possible combinations.
Thus the extremal dependence between the components of $X$ may have a complicated structure with both asymptotic dependence and independence present. In dimension $d=3$, Figure~\ref{fig:simplex} shows the $\ell_1$-simplex $\mathbb S^2_+$ and its intersections with the 7 different faces $\EE_I$, together with the observations of the extremal angle $\Theta$ for three stations of the river data set from Section~\ref{river}.

If the exponent measure $\Lambda$ is absolutely continuous with respect to $d$-dimensional Lebesgue measure we denote its density by $\lambda$. In this case, both the max-stable distribution $Z$ and the multivariate Pareto distribution $Y$ also possess densities. More generally, $Z$ has a density if and only if $\Lambda$ has a density on each face $\EE_I$, $I \subset V$ \citep{dom2016b}.

\subsection{Empirical estimation}\label{sec:stat}

The exponent measure $\Lambda$ can be estimated empirically. Let $X^{(1)},\dots, X^{(n)}$ be independent observations of the random vector $X$.
Using Equation~\ref{eq:mu} with $t = n/k$, we define an empirical
estimator of $\Lambda_z = \Lambda(\EE \setminus [0,z])$, $z \in \EE$, by
\begin{align}\label{emp_lambda}
  \widehat \Lambda_z = \frac1k \sum_{i=1}^n \einsfun\left\{ \widehat F_1(X^{(i)}_{1}) > 1 - \frac{k}{nz_1} \text{ or } \dots \text{ or } \widehat F_d(X^{(i)}_{d}) > 1 - \frac{k}{nz_d}\right\},
\end{align}
where $k = k(n)$ can be interpreted as the number of exceedances.
Here, the application of the empirical distribution function $\widehat F_j$ corresponds to the standardization explained in Section~\ref{sec:mrv}.
This estimator is closely related to the empirical estimator of the stable tail dependence function, whose asymptotic behavior is well studied \citep[see][]{deh2006a}. Under the standard assumption that $k \to \infty$ and $k/n \to 0$, $\widehat \Lambda_{z}$ is a consistent estimator of $\Lambda_z$, and under appropriate second-order conditions it is asymptotically normal even in a functional sense as a process indexed by $z \in \EE$ \citep[see][]{H1992,DH1998,EKS2012,BSV2014}. 
The estimators in Figure~\ref{fig:chi_plot} are obtained as $\widehat \chi_{ij}(q) = 2- \widehat \Lambda_{(1,1)}$ with $q=1-k/n$ for a range of values for $k$.

Using a similar counting approach as in Equation~\ref{emp_lambda}, we can define the empirical version $\widehat H$ of the angular distribution $H$ of the extremal angles $\Theta$ based on Equation~\ref{eq:mrv}, and asymptotic theory for this estimator can be derived.

\section{Classical models and their limitations}
\label{sec:classical}

In recent decades, there has been active research on the construction of
statistical models for multivariate extremes. In this section we give a brief overview of the literature and mention some of the models that appear in later parts of the review. We also
describe the limitations of these classical approaches when facing more complex data sets in
higher dimensions. One of the simplest models is the max-linear model.
\begin{example}[Max-linear model]\label{max_linear}
  Let $\eps_i$, $i =1,\dots, p$ be independent standard Fr\'echet variables and $A=(a_{ij})$
  be a $d\times p$ matrix of non-negative coefficients. Define a $d$-dimensional random vector $Z$ with entries
\begin{align}  \label{ML_model}
  Z_i=\max(a_{i1}\eps_1,\dots, a_{ip}\eps_p),\quad i=1,\dots, d.
\end{align}
We assume that the rows of $A$ sum up to~$1$, ensuring that all $Z_i$ have standard Fr\'echet distributions. The max-linear model is max-stable with exponent measure $\Lambda$ supported on $p$ rays specified by the columns of $A$, or in other words, the angle $\Theta$ takes $p$ possible values $a_{\cdot j}/\|a_{\cdot j}\|$ with probabilities proportional to $\|a_{\cdot j}\|$. 
\end{example}
Every max-stable distribution whose angular measure $H$ concentrates on finitely many points corresponds to a max-linear model; see \cite{yue2014} for details. For applications, max-linear models are often too simplistic but they may provide a first approximation of the data and are useful tools to illustrate statistical methods. For instance, it is easy to construct a max-linear model whose exponent measure has support on any combination of faces of $\mathcal E$.
More realistic parametric model classes are often specified in terms of the exponent measure density $\lambda$.
\begin{example}[Logistic distribution]\label{ex_log}
  The $d$-dimensional extremal logistic distribution with parameter $\theta\in(0,1)$ has exponent measure density
  \begin{align}\label{dens_log}
    \lambda(y) =  \left(\sum_{i=1}^d y_i^{-1/\theta}\right)^{\theta-d}\prod_{i=1}^{d-1}\left(\frac{i}{\theta}-1\right) \prod_{i=1}^{d} y_i^{-1/\theta-1}, \quad  y \in \mathcal E.
  \end{align}
  The strength of dependence between all components ranges from complete dependence for $\theta\to 0$ to independence for $\theta \to 1$. 
\end{example}
The logistic distribution is symmetric and has only one parameter $\theta$, independently of the dimension. The asymmetric logistic distribution \citep{taw1988a} is an extension that is more flexible, but at the price of an order of $2^d$ parameters in dimension $d$.
A parametric family with good control of extremal dependence between any pair of components is the
distribution introduced in \citet{Husler1989}. For this, and many other reasons, it can be seen as the Gaussian distribution for asymptotically dependent extremes. 

\begin{example}[\HR{} distribution]\label{ex:HR}
This distribution is parameterized by a variogram matrix $\Gamma = (\Gamma_{ij})_{i,j\in V}$ and its exponent measure density can be written for any $m\in V$ as \citep[see][]{Engelke2015}
\begin{align}\label{eq:fYHR}
  \lambda(y)
  &= y_m^{-2}\prod_{i\neq m} y_i^{-1} \phi_{d-1}\left( \log(y_{- m}/y_m) + \Gamma_{-m, m}/2; \Sigma^{(m)}\right), \quad  y \in \mathcal E,
\end{align}
where $\phi_p(\cdot; \Sigma)$ is the density of a centered $p$-dimensional normal distribution with covariance matrix $\Sigma$, the notation $-m$ in a vector or matrix means omission of the $m$th component, and 
\begin{align}\label{sigma_k}
  \Sigma^{(m)}  =\frac{1}{2} \{\Gamma_{im}+\Gamma_{jm}-\Gamma_{ij}\}_{i,j\in V \setminus\{m\}} \in\mathbb R^{(d-1)\times (d-1)}.
\end{align}
The strength of dependence between the $i$th and $j$th components is parameterized by $\Gamma_{ij}$, ranging from complete dependence for $\Gamma_{ij}=0$ to independence for $\Gamma_{ij}=\infty$. 
\end{example}
There are many further multivariate models such as the Dirichlet mixture model in \cite{BoldiDavison}
or the pairwise beta distribution in \cite{CooleyDavisNaveau}; we refer to \citet{seg2010} for a detailed overview.

All of these model classes become restrictive in higher dimensions, either because of a lack of flexibility or a rapid increase in the number of parameters. Parsimonious extreme value models in large dimensions have been developed for spatial applications, where the vector $X$ is recorded at $d$ locations
$t_1,\dots, t_d\in \mathbb R^D$ in space. Following ideas from geostatistics \citep[see][]{wac2013}, extremal dependence is then parameterized in terms of distances $\|t_i - t_j\|$, which drastically reduces the number of required parameters.
Brown--Resnick processes \citep{bro1977, kab2009} for instance, are the extension of H\"usler--Reiss distributions to random fields and are widely used models for spatial rare events. Other models
for such max-stable processes have been introduced in \citet{sch2002, opi2013} and \citet{rei2012}, in \citet{dav2013} and \citet{dav2015} for the spatio-temporal setting, and in \citet{asadi2015extremes} for river networks.
Without domain knowledge such as the spatial locations of gauging stations, these models can no longer be applied. For general multivariate data, there are some approaches to define low-dimensional parametric representations through copulas \citep{aas2009,lee2018}, graphical constructions \citep{Hitz2015} and elliptical distributions~\citep{haug2009dimension}, or as ensembles of trees \citep{yu2017}. 

Statistical inference for multivariate extreme value models is challenging, and the related literature is vast. Maximum likelihood estimation is commonly used \citep[e.g,][]{Engelke2015, wad2013, thi2015} but can be computationally demanding due to censoring that is applied to non-extreme components \citep[see][]{led1997}. Alternative methods include pairwise likelihood \citep[e.g,][]{var2011, pad2010}, $M$-estimation \citep[e.g,][]{ein2012, ein2016} and proper scoring rules \citep{def2018}. Exact simulation of these models, both conditional and unconditional, has also been studied \citep[][]{dom2013, die2015, dom2016}.

While spatial models have few parameters even in high dimensions, they possess some major limitations. On the one hand, they require prior domain knowledge on the
spatial locations of gauging stations and rely on the strong assumption of stationarity in space. It is not possible to learn the underlying structure from the data. On the other hand, even though these models have few parameters, their distributions do generally not exhibit any sparsity properties in a probabilistic sense, such as conditional independence patterns or support on low-dimensional sub-spaces. This means that statistical inference does not simplify and likelihood inference is limited to fairly moderate dimensions \citep{thi2015, dom2016a, hus2019}.

In the next sections we present a new line of research providing alternatives to these classical methods. They learn sparse structures and low-dimensional representations in multivariate extreme values in a data driven way and do not require additional domain knowledge or stationarity assumptions.

\section{Adaptation of unsupervised learning methods}\label{sec:adapt}
Clustering and principal component analysis are two of the standard methods in multivariate analysis~\citep[see][]{anderson_book}. They are both tools to detect lower-dimensional representations of the data. For extremes, there is recent work that adapts these tools to find structures in multivariate tails assuming the following notion of sparsity.
\begin{itemize}
	\item[(S1)]
          The dimension of the support of the exponent measure $\Lambda$ is much smaller than~$d$.
        \end{itemize}
In other words, the exponent measure can be expressed via a low-dimensional object. In this section we describe the expanding literature in this field.

\subsection{Clustering approaches}
\label{sec:clustering}
Centroid-based clustering aims at finding the set of $p$ points $c_1,\ldots, c_p \in \R^d$, called cluster centers, that minimize the cost
\begin{equation}\label{eq:clustering}
  \E \min_{j=1,\dots,p}\varrho(\Theta,c_j),
\end{equation}
where $\Theta\in \R^d$ is a random object of interest and $\varrho$ is a given distance or dissimilarity function.
In the setting of extremes the main focus is on the case where $\Theta$ is the extremal angle appearing in the decomposition of the exponent measure $\Lambda$ in Equation~\ref{spectral_decomp} with distribution $H$.
The above optimization problem is computationally hard and usually heuristic algorithms are used that exhibit fast convergence to a local optimum.
Clustering is mainly an exploratory tool which may lead to dimension reduction in two ways. Firstly, the associated cost may become small for a moderate number of clusters~$p$. This happens when the angular distribution concentrates at a small number of points in $\mathbb S^{d-1}_+$, thereby hinting at a max-linear model as in Example~\ref{max_linear} and a sparse representation as in assumption (S1). Secondly, all of the cluster centers may have multiple small entries indicating that $H$ puts mass only on some faces of $\mathbb S^{d-1}_+$; see Section~\ref{sec:faces} for this notion of sparsity.

\cite{cha2015} and \cite{jan2019} propose clustering the angle $\Theta$ using the spherical $k$-means procedure of~\cite{dhi2001}, which ensures that cluster centers also belong to the simplex~$\mathbb S^{d-1}_+$.  \cite{jan2019} employ the angular dissimilarity
\begin{align}\label{ang_diss}
  \varrho(x,y)=1-\cos(x,y)=1-\frac{x^\top y}{\|x\|_2\|y\|_2},
\end{align}
which is independent of the norm used to define~$\Theta$. They establish a consistency result showing that cluster centers obtained from the empirical distribution of angles $\widehat H$ (see Section~\ref{sec:stat}) converge to the cluster centers of the true angular distribution $H$. It is noted, that this empirical approximation does depend on the choice of norm.
 They further investigate an application of this clustering method to inference for max-linear models defined in Equation~\ref{ML_model}, where the angular distribution $H$ concentrates on $p$ points $a_{\cdot j}/\|a_{\cdot j}\|\in\mathbb S^{d-1}_+$, $j=1,\dots, p$. This method provides estimates of the parameter vectors $a_{\cdot j}$ that are competitive or even superior to other estimation methods considered by~\cite{yue2014} and \cite{ein2016, ein2018}. Finally, \citet{jan2019} suggest using the cluster centers as `prototypes of directions of extremal events', an idea that we adopt in our flood application in Section~\ref{conc_application}.

Clustering in the context of extremes is also considered in \cite{bernard2013clustering}, however of a very different nature. They suggest grouping the components of $X$ using a certain extremal dissimilarity similar to the pairwise tail dependence coefficients $\chi_{ij}$ as the distance between two components $X_i$ and $X_j$; see also \citet{Saunders2019} for an application of this method to rainfall extremes.

\subsection{Principal component analysis}\label{sec:PCA}
Principal component analysis (PCA) is a classical method of multivariate analysis to reduce the dimension of a random vector $W\in\R^d$ while capturing most of its variability.
PCA identifies the linear subspace  $\mathcal S^*\subset \R^d$ of a given dimension $p<d$ so that the $\ell_2$-distance 
\begin{equation}\label{eq:PCArisk}\E\|\Pi_{\mathcal S^*} W-W\|_2^2\end{equation}
between $W$ and its projection $\Pi_{\mathcal S^*}W$ onto $\mathcal S^*$ is minimal~\citep[see][]{seber_book}, and thus $\Pi_{\mathcal S^*}W$ can be seen as the best $p$-dimensional approximation of $W$.
Fundamental to PCA are the orthonormal eigenvectors $v_1,\ldots, v_d\in \mathbb R^d$ of the positive semi-definite matrix $\Sigma=\e (WW^\top)$, ordered according to the respective eigenvalues $\lambda_1\geq \cdots\geq \lambda_d\geq 0$. 
The linear span of the first $p$ eigenvectors $v_1,\ldots, v_p$ yields the desired $S^*$, 
whereas the best $p$-dimensional approximation of $W$ is obtained by
summing up the respective orthogonal projections, called principal components,
\[\Pi_{\mathcal S^*}W=\Pi_{v_1}W+\cdots+\Pi_{v_p}W.\]
Importantly, PCA results in an iterative procedure, often called reconstruction of $W$, where the principal components are added until the approximation error in Equation~\ref{eq:PCArisk} drops below a certain threshold. 
For a zero mean vector $W$ the optimization criterion in Equation~\ref{eq:PCArisk} is equivalent to maximizing the variance of the projection~$\Pi_{\mathcal S^*}W$.
For statistical properties of PCA and theoretical guarantees  we refer to~\cite{blanchard2007statistical}.

In the present setting one is interested, loosely speaking, in discovering a lower-dimensional linear subspace explaining most of the extreme behavior. 
\cite{coo2019} and \cite{dre2019} consider the extremal angle $\Theta$ with distribution $H$ and
the respective matrix 
\begin{equation}\label{eq:sigmaPCA}
  \Sigma=\E(\Theta\Theta^\top)=\lim_{t\to \infty}\E\left(\left.\frac{XX^\top}{\|X\|^2}\, \right|\,\|X\|>t\right),
\end{equation}
which was introduced by \cite{lar2012} in the bivariate case. 
As explained above, the aim is to identify the optimal $p$-dimensional linear space $\mathcal S^*\subset \R^d$ for $\Theta$.
In applications, the distribution of $\Theta$ is replaced by its empirical estimate $\widehat H$ (see Section~\ref{sec:stat}). 
\cite{dre2019} show that, as the sample size $n\to \infty$, the corresponding optimal $p$-dimensional linear spaces converge in probability to the true $\mathcal S^*$, provided the latter is unique, with respect to the metric 
$$\varrho(\mathcal S,\mathcal S')=\sup_{\theta\in\mathbb S_+^{d-1}}\| \Pi_\mathcal S \theta-\Pi_{\mathcal S'} \theta\|_2.
$$
Importantly, the projection $\Pi_{\mathcal S^*}\Theta$  does not necessarily belong to $\mathbb S_+^{d-1}$.
Nevertheless, if the linear span of the support of~$H$ has dimension $p$ then this linear span is the optimal~$\mathcal S^*$ and the loss in Equation~\ref{eq:PCArisk} is zero. 
Intuitively, slight deviations from this assumption would lead to a projection in a neighborhood of $\mathbb S_+^{d-1}$, which then can be normalized/shifted appropriately. If the loss in Equation~\ref{eq:PCArisk} is not negligible then this method may produce a sub-optimal approximation of~$\Theta$ in the given dimension~$p$.

If the marginals of $X$ are standardized so that the second moments exist, the above PCA for the angle~$\Theta$ is equivalent to PCA for the limit distribution of $(X/t\,|\,\|X\|>t)$, because the radial component becomes independent of the angle.
\cite{coo2019} follow this interpretation and suggest a way to reconstruct extreme scenarios of the original vector~$X$. The problem that projections on $\mathcal S^*$ may not belong to the domain of interest does however not disappear, and projections of $X$ onto $v_i$ may have negative entries. To remedy this, the authors propose projecting $t^{-1}(X)$ on $\mathcal S^*$ for some bijection $t:\R\mapsto \R_+$ behaving as identity for large arguments, and then applying $t$ to get back to the positive orthant. It is noted that the choice of the mapping $t$ and the choice of the marginal distributions are somewhat arbitrary, but may have a major influence on the resulting approximation.

\cite{cha2015} suggests using a technique called principal nested spheres developed by~\cite{jung12}.
Firstly, $\Theta$ is renormalized to lie on the $\ell_2$-sphere, and then it is projected on a sub-sphere of dimension $d-2$ formed by intersecting the original sphere with a hyperplane. Using numerical optimization, this sub-sphere is chosen to minimize the $\ell_2$-norm of the empirical geodesic distance to~$\Theta$. The procedure is iterated until a certain loss exceeds a pre-defined threshold, thus resulting in a greedy search for a lower-dimensional approximation. 
In the setting of extreme angles, \cite{cha2015} motivates the projection on sub-spheres by the problem of discovering mass of $H$ on the faces; see also Section~\ref{sec:faces}. 
Unlike PCA the final result may not be optimal and the computational effort is considerably larger. Furthermore, the approximation lies on the sphere but may not be in the positive orthant.

\subsection{Application to flood risk}\label{unsupervised_application}

We conclude this section with an application to the river flow data set discussed in Section~\ref{river}.
As suggested by~\cite{dre2019} and~\cite{coo2019} we apply PCA to the matrix $\Sigma$ in Equation~\ref{eq:sigmaPCA}. We use the empirical $90\%$-quantile of the radius $\|X\|$ as the threshold and compute the estimate $\widehat \Sigma$ based on the approximate extremal angles of the $k=202$ exceedances (see Section~\ref{sec:stat}). We do not use temporal declustering since \citet{zou2019} show that a larger but possibly dependent data set can decrease the asymptotic estimation error.
The $\ell_2$-norm was used so that the eigenvalues sum up to~1, but the results for the $\ell_1$-norm are similar. 

\begin{figure}
    \centering
    \begin{subfigure}[b]{0.49\textwidth}
        \includegraphics[trim=1cm 1.5cm 0cm 0cm, width=.95\textwidth]{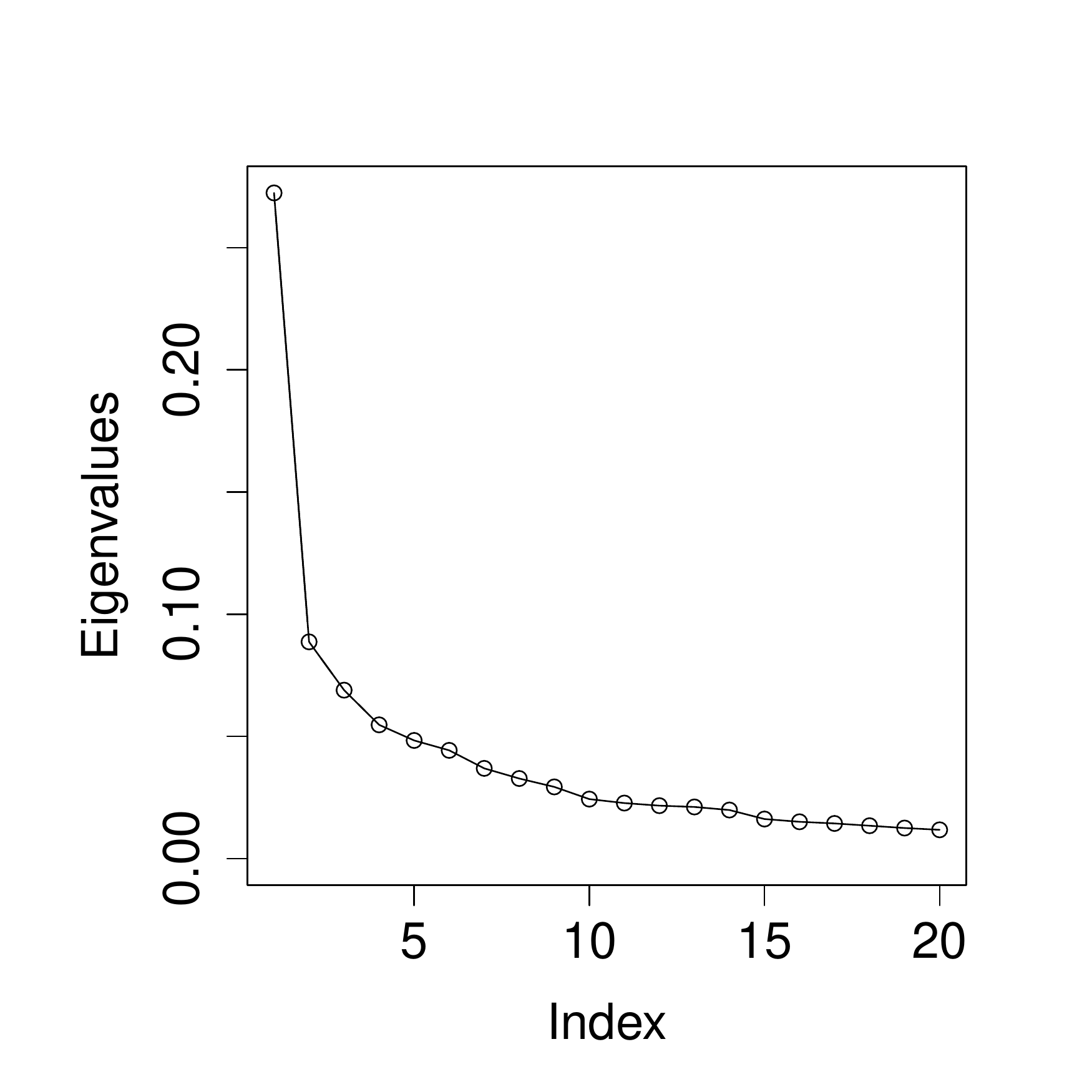}
    \end{subfigure}
    \begin{subfigure}[b]{0.49\textwidth}
        \includegraphics[width=\textwidth]{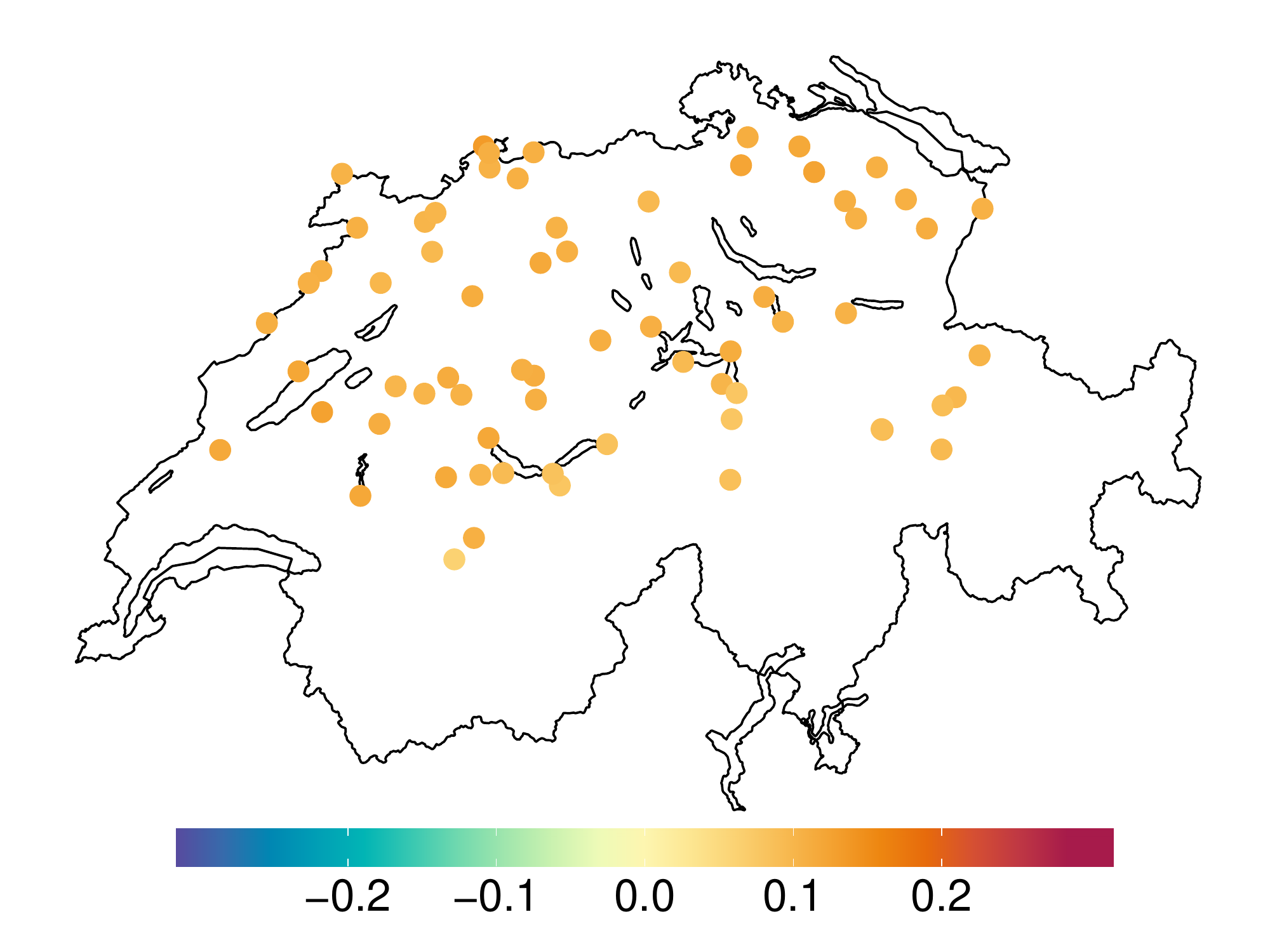}
    \end{subfigure}
        \begin{subfigure}[b]{0.49\textwidth}
        \includegraphics[width=\textwidth]{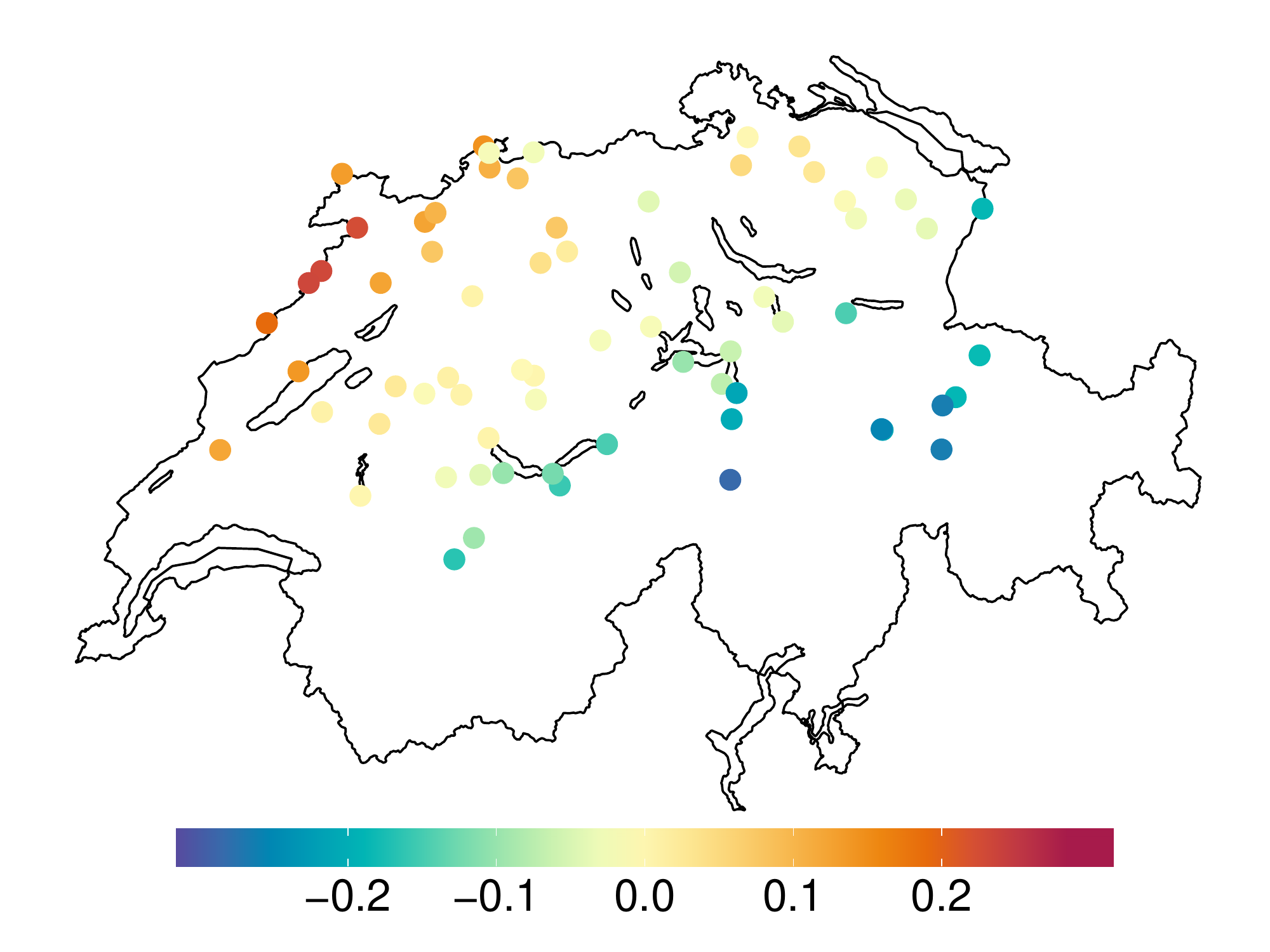}
    \end{subfigure}
        \begin{subfigure}[b]{0.49\textwidth}
        \includegraphics[width=\textwidth]{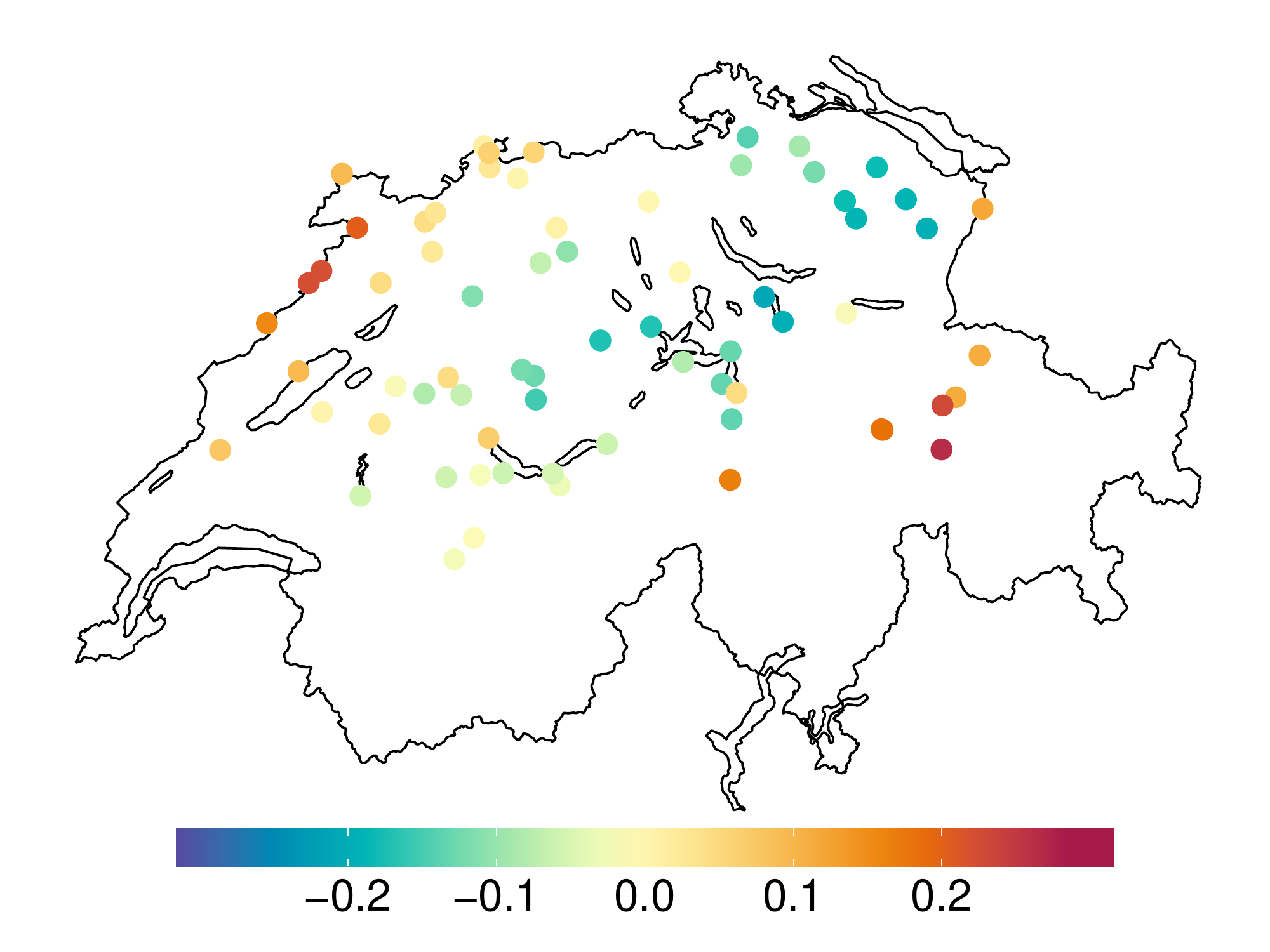}
    \end{subfigure}
        \caption{Scree plot of eigenvalues (top left) and the first three eigenvectors plotted at the corresponding geographical locations of the stations.}\label{fig:pca}
\end{figure}

Even though we do not use information on the geographical locations, the first three eigenvectors
shown in Figure~\ref{fig:pca} exhibit a clear spatial pattern. While the leading eigenvector points to the center of the simplex, the second eigenvector shows a linear trend from southeast to northwest. The third eigenvector exhibits a slightly more complicated spatial pattern. Figure~\ref{fig:pca} also shows a scree plot of the eigenvalues which quickly flattens. It is noted, however, that for $p=3$ the mean squared loss in Equation~\ref{eq:PCArisk} evaluates to $1-\sum_{i=1}^3\lambda_i=0.57$, which yields root mean squared error of about~$0.75$.
The first three principal components thus explain $25\%$ of the extremal dependence, while the first 20 explain about~$65\%$.

\section{Concomitant extremes}\label{sec:faces}
The identification of groups of variables that can be large simultaneously is one of the basic questions in multivariate extremes. Such groups $I\subset V$ correspond to $\chi_I>0$.
The stronger condition $\Lambda(\EE_I)>0$ asserts that the components indexed by $I$ can be extreme while the others are much smaller (see Section~\ref{sec:props}).
In this respect, multivariate extremes are different from classical multivariate analysis where restriction of some variables to zero normally is of little interest.

\begin{example}\label{ex:simplex}
  Figure~\ref{fig:simplex} shows the extremal angles corresponding to the sub-group $\{1,56,57\}$ of stations in the river application. In this case, visually, it seems that $\Lambda$ puts mass on the faces $\EE_{\{1\}},\EE_{\{2,3\}}$ and possibly $\EE_{\{1,2,3\}}$. Note that station $1$ is far apart from stations $56$ and $57$, whereas the latter two are close-by; see Figures~\ref{fig:basin} and~\ref{fig:chi_plot}. This means that floods can either occur at all stations simultaneously, possibly due to a larges-scale precipitation event, or separately either at station $1$ or at both stations $56$ and $57$ due to heavy localized rain. 
\end{example}

In principle, it is possible that all combinations of extreme components may arise
and $\Lambda$ has mass on all $2^d-1$ faces. This situation, however, is rather
unlikely in data applications, since we always expect some structure between the
variables as in Example~\ref{ex:simplex}, and simply since the number of exceedances $k(n)$ is usually much smaller than $2^d-1$. Moreover, current statistical models are not flexible enough to jointly model
complex dependence structures on all possible faces. They are mostly designed to separately model different subsets of variables which can be concomitantly extreme.
If we knew the relevant faces and the corresponding probabilities, a sensible modeling strategy would be to fit models
on these faces separately and to combine them as a mixture model.

In order to make this feasible, one may assume a notion of sparsity in terms of the
number and dimension of the faces of $\EE$ charged with mass \citep[see][]{goi2017}.
\begin{itemize}
	\item[(S2.a)]
	There is only a small number of groups of variables in $(X_j: j \in V)$ that
	can be concomitantly extreme, that is,
	$$\left|\{ I\subset V : \Lambda(\EE_I) > 0 \} \right| \ll 2^d - 1.$$
	\item[(S2.b)]
	Each of these groups contains only a small number of variables, that is,
	$$\max\{|I|: \Lambda(\EE_I) > 0\} \ll d.$$
      \end{itemize}
The sparsity notion (S2.a) is the most crucial since it limits the number of components in a mixture model. 
If, in addition, the second notion (S2.b) holds, then each of the sub-models
is low-dimensional and particularly simple. 
As we will see in Section~\ref{sec:markov}, a different notion
of sparsity for densities on the cones $\EE_I$ can be defined that allows to model
simultaneous extremes for large~$|I|$.

\subsection{Detecting faces with $\Lambda$-mass}\label{sec:mumass}
Recall the set $\LL$ bounded away from the origin and consider its partition into 
\[\LL_I=\{x\in \EE_I: \|x\|_\infty\geq 1\}.\]
We are interested in identifying subsets $I$ such that  $\Lambda(\LL_I)>0$, as well as the respective masses  characterizing the weights in the mixture model.
The main difficulty in estimation of these masses is that the convergence in Equation~\ref{eq:mu} does only hold for $\Lambda$-continuous sets, excluding sets $\LL_I$ charged with mass.
We therefore cannot simply rely on empirical estimates of the left-hand side of Equation~\ref{eq:mu}, because even for large $t>0$, there will typically be no observation of $X/t$ falling in $\LL_I$ since none of the components is exactly zero.

\begin{figure}
\centering
\begin{tikzpicture}[scale=0.8]
\draw [dashed,red_new,domain=0:3,opacity=0.6] plot (\x, {0.1*\x^0.85});
\fill[color=red_new,domain=3:8,opacity=0.6] plot (\x, {0.13*\x^0.85})--(8,0)--(3,0);
\fill[color=blue_new,opacity=0.5] (3,0.5) -- (8,0.5)--(8,0)--(3,0);
\fill[color=green_new, opacity=0.35](3,0)--(8,5)--(8,0);

\fill[color=red_new,opacity=0.6, domain=(3^(0.85)*0.13):(8^(0.85)*0.13)] plot (\x, {(\x / 0.13)^(1/0.85)})--(0,8)--(0,3);
\fill[color=blue_new,opacity=0.5] (0.5,3) -- (0.5,8)--(0,8)--(0,3);
\fill[color=green_new, opacity=0.35](0,3)--(5,8)--(0,8);

\draw[thick,->] (0,0) -- (8.2,0);
\draw[thick,->] (0,0) -- (0,8.2);
\node[below] at (7.5,0) {$X_1/t$};
\node[left] at (0,7.5) {$X_2/t$};
\node[left] at (0,3) {$1$};
\node[below] at (3,0) {$1$};
\node[color=red_new] at (2,0.5) {$t^{\delta-1}(\cdot)^\delta$};
\draw[dashed, blue_new,opacity=0.5] (0.5,0) -- (0.5,3);
\node[below,color=blue_new] at (0.5,0) {$\varepsilon$};
\node[below,color=white] at (0,-.8) {};
\end{tikzpicture}
\caption{The regions considered by~\cite{goi2017} (blue), \cite{sim2018} (red) and \cite{mey2019} (green) 
to study the masses on faces $\EE_{\{1\}}$ (bottom right) and $\EE_{\{2\}}$ (top left) in the bivariate case.}
\label{fig:masses}
\end{figure}
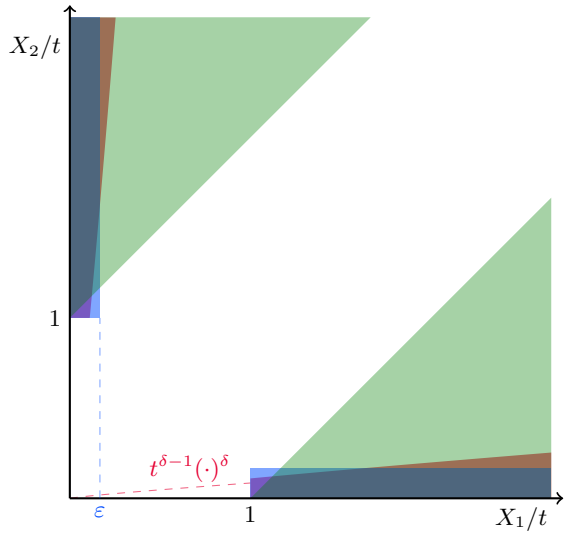

To circumvent this difficulty,  \cite{goi2016,
  goi2017}, propose to partition $\LL$ into sets
\begin{align}\label{def_L}
  \LL_I^\varepsilon=\{x\in \LL: x_i>\varepsilon\,\forall i\in I, x_j \leq \varepsilon\,\forall j\notin I\}
\end{align}
for some small $\varepsilon>0$, which they call $\varepsilon$-thickened rectangles; see the blue regions in Figure~\ref{fig:masses}.
In this setting Equation~\ref{eq:mu} is valid, and we can approximate $\Lambda(\LL_I^\varepsilon)$ by $t\p(X/t\in \LL_I^\varepsilon)$ or rather its empirical estimate for some large threshold~$t$ (see Section~\ref{sec:stat}).
Since $\Lambda(\LL_I^\varepsilon)$ converges to $\Lambda(\LL_I)$ as $\varepsilon \to 0$, the authors argue that if $\varepsilon$ is chosen small enough then
essentially only mass that corresponds to the set $\LL_I$ will enter the estimate. If this empirical mass is larger than a threshold $u$, then the $I$th face is identified to have positive $\Lambda$-mass.
Both thresholds $\varepsilon$ and $u$ are tuning parameters. One issue with this approach is that it may be that $\Lambda(\LL_I)=0$ even though non-negligible mass  $\Lambda(\LL_I^\varepsilon) > 0$ in the corresponding $\varepsilon$-thickened rectangle is detected. Too many groups of concomitant extremes may thus be identified, which is confirmed by simulation studies in \cite{chi2017} and \cite{sim2018}.  As discussed above, this is a serious issue countering the sparsity assumption (S2.a). A larger threshold~$u$ for the mass may not be an option since then too many exceedances are ignored.

In order to cope with this problem  \cite{sim2018} suggest replacing the upper bound in the definition of $\LL^\varepsilon_I$ in Equation~\ref{def_L} by a more flexible threshold-dependent constraint $\varepsilon_t$ with $\varepsilon_t \to 0$ as $t\to \infty$. This still allows that $\lim_{t\to\infty}t\p(X/t\in \LL_I^{\varepsilon_t}) > 0$.
More precisely, their modeling strategy is to consider
\begin{align}\label{sim_model}
\mathbb P\left( \min_{i\in I}(X_i/t)>1, \max_{i\notin I}(X_i/t)\leq  t^{\delta-1}\Big(\min_{i\in I} (X_i/t)\Big)^\delta\right),
\end{align}
for some $\delta\in[0,1)$,
and to assume that this probability is regularly varying as $t\to\infty$; see the red regions in Figure~\ref{fig:masses}.
Note that if the probability in Equation~\ref{sim_model} is of order $1/t$ then $\EE_I$ must have a positive mass.
For a fixed subset~$I$, \cite{sim2018} use the Hill estimator \citep{hil1975} to fit the model and then extrapolate this probability for large values of $t$.
They propose to identify $I$th face to have positive $\Lambda$-mass if the corresponding approximation is
larger than a suitable threshold $u$, similarly to \cite{goi2017}.
Finally, the choice of $\delta\in [0,1)$ requires a subtle trade-off between the significance level and the power of detecting mass on $\EE_I$. If $\delta$ is too close to one, then this procedure runs into the same issues as the method of~\citet{goi2017}.
On the other hand, if $\delta$ is too small, then it can happen that no mass is detected even though $\Lambda(\EE_I)>0$.

\begin{remark}\label{rem:comlexity}
	Importantly, the above procedures do not necessarily require processing all $2^d-1$ faces. 
	Note that we can not identify more faces with mass than there are exceedances. Thus instead of cycling over the faces we can go over all exceedances, whose number is fairly small by definition.
\end{remark}	

\subsection{Detecting maximal faces with $\Lambda$-mass}\label{sec:maxfaces}

Recovering all faces $\EE_I$ with positive mass is a difficult problem. Moreover, it does not lead to a sparse representation when the mass is spread over a large number of faces.
For the data set on river discharges studied in \cite{chi2017}, it was found that many of the detected groups of variables differ from each other only by a single or two elements. Practically speaking, this means that several distinct extreme events have impacted almost the same set of stations.
This motivates gathering such groups into a single one and a natural approach is to look at maximal sets.
More precisely, the aim is to identify groups $I$ with $\Lambda(\EE_I)>0$ such that $\Lambda(\EE_J)=0$ for all $J\supsetneq I$.   
As noted by~\cite{chi2017} this is the same as looking for the maximal sets with $\chi_I>0$.
Indeed, the latter implies that the variables indexed by~$I$ can be simultaneously extreme, and by maximality of $I$ no further variables can be added.

\citet{chi2017} propose to use the Apriori algorithm \citep{agr1994} for frequent item set mining with a novel stopping criterion.
This algorithm results in a bottom-up approach starting with singletons and at each step enlarging all the groups by one element if there is sufficient evidence that all the components can be concomitantly extreme.
They use a threshold-based stopping criterion involving the empirical estimate of a conditional version of $\chi_I$.
This conditional tail dependence coefficient is taken to avoid the problem that $\chi_I$ decreases as the sets grow; see the comment following Equation~\ref{eq:chi_eta_ineq}.
This work was extended by \cite{chi2019} proposing three other stopping criteria based on formal hypothesis testing.
In particular, they consider testing whether the residual tail dependence
coefficient in Equation~\ref{eq:eta} satisfies $\eta_I = 1$, which, under a weak assumption, is equivalent to $\chi_I > 0$. 
They further derive asymptotic results for controlling the type-I error of this test, and show that it has a better performance in simulation studies. 
Such tests are applied to every sub-face of a maximal face with mass, which leads to multiple testing problems and potentially long running times. 
The Apriori algorithm is thus efficient only when (S2.b) holds true, whereas
	it has to pass through all $2^d-1$ subsets if $\Lambda(\EE_V)>0$.

\cite{mey2019} suggest another approach for recovering the maximal sets with mass.
For all observations where $\| X  \|_1 > t$, for a large $t>0$, instead of the usual projection $X/\|X\|_1$, they use the Euclidean projection $\pi_1(X/t)$ \citep[see][]{duchi2008efficient} onto the positive $\ell_1$-sphere $\mathbb S^{d-1}_+$. The projection $\pi_1(x)$, $x \in \mathbb R^d_+$, is the point on $\mathbb S^{d-1}_+$ that minimizes the $\ell_2$-distance to $x$.
The limiting mass on the the $I$th face of the $\ell_1$-sphere is
\begin{align}\label{pi_mass}
m_I = \lim_{t\to\infty} \mathbb P\left(\pi_1(X/t) \in \EE_I    \mid \| X  \|_1 > t \right),
\end{align}
see the green regions in Figure~\ref{fig:masses}.
The geometry of the Euclidean projection has the effect that possibly more mass is projected on sub-faces of $\mathbb S^{d-1}_+$ than the spectral measure actually has (see Equation~\ref{eq:mrv}). Importantly, the maximal sets in $\{I\subset V: m_I>0\}$ coincide with the maximal sets having positive $\Lambda$-mass.
Empirical estimates $\widehat m_I$ of $m_I$ are used in \cite{mey2019} to find these maximal sets. 

We conclude this section by noting that \cite{lehtomaa2019asymptotic} study a somewhat related problem of estimating the support of the extremal angle~$\Theta$.

\begin{remark}\label{rem:apriori}
	The methods above for detecting maximal faces with $\Lambda$-mass have the clear limitation that if
	there is mass on a high-dimensional face, say $\Lambda(\EE_V)>0$, then no other face $\EE_I$ with
	$I\subsetneq V$ can be detected, even if $\Lambda(\EE_I) > 0$. 
	These
	methods must therefore assume the sparsity notion (S2.b), since otherwise too much information
	is lost. 
\end{remark}

\subsection{Application to flood risk}\label{conc_application}
We reconsider our flood risk application from Section~\ref{river} and aim to identify the groups of variables which can be concomitantly extreme.
We do not attempt to fine-tune each of the methods, but rather to illustrate the main ideas.

As has already been observed in the literature~\citep{chi2017,chi2019}, the truncation method of~\cite{goi2017} yields a very large number of faces, most of which have a single associated extremal observation. The only faces with more observations are typically of dimension close to~$d$. 

\begin{figure}
	\includegraphics[trim=0 1cm 0 0, width=\textwidth]{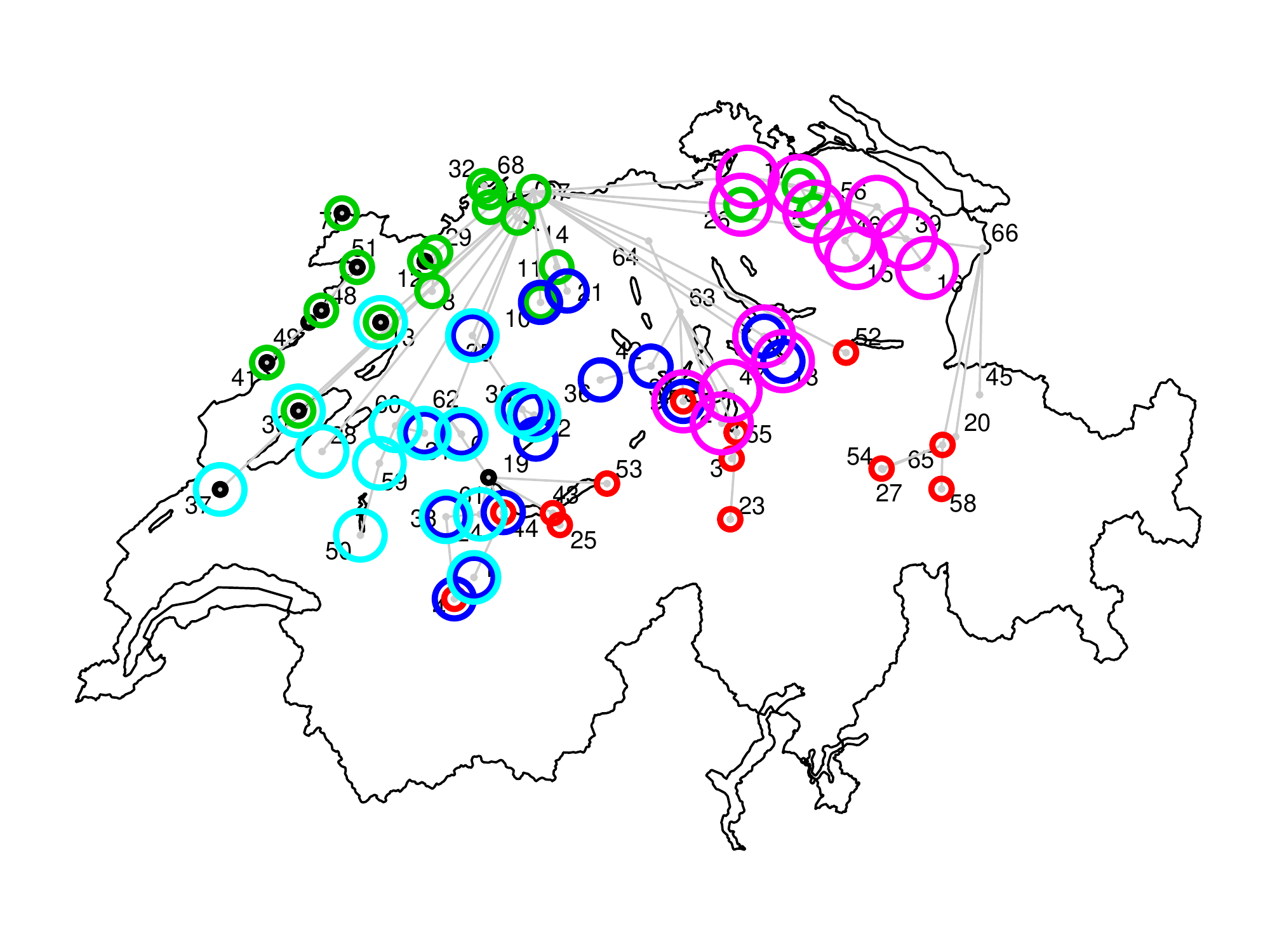}
				\caption{The 6 faces with more than 20 exceedances obtained by clustering angles into 10 groups and thresholding the centers at $0.02$; the stations associated to each face have the same color and circle radius. The gray lines show the flow connections between the stations.}
	\label{fig:clusters}
\end{figure}

We therefore follow the idea of~\cite{jan2019}, explained in Section~\ref{sec:clustering}, and use clustering to find the prototypes of groups of concomitant extremes, to which we then apply thresholding as in~\cite{goi2017}.  Firstly, we obtain samples of the extremal angle $\Theta$ for the $\ell_1$-norm (see Section~\ref{unsupervised_application}) and then cluster them into  $p=10$ groups using the angular dissimilarity in Equation~\ref{ang_diss}. To each cluster center $c_j\in\mathbb S^{d-1}_+$, $j=1,\dots, p$, we associate the face $I_j=\{i\in V:c_{ij}>0.02\}$, where the cut-point $0.02$ is a tuning parameter regulating the number of components in each group. The resulting faces are all of  moderate dimensions; see Table~\ref{tab:clustering}.
Figure~\ref{fig:clusters} shows the faces with more than $20$ associated exceedances.
The cluster $\#10$ has the highest number of exceedances and the corresponding group $I_{10}=\{2,  5,  9, 15, 16, 17, 18, 26, 34, 39, 46, 47, 56, 57\}$ (in magenta color) will be modeled after some minor changes in Section~\ref{sec:markov}.
Interestingly, the components of each face are grouped geographically even though no such information is used. We note that clustering with a sightly different number of clusters produces similar results. One can also increase the number of clusters, which eventually leads to associating each observed angle with a face~\cite[see][]{goi2017}.

\begin{table}
\centering
\caption{The 10 different clusters with the dimensions of the associated faces and the number of exceedances.}
\label{tab:clustering}
\begin{tabular}{c|c|c|c|c|c|c|c|c|c|c||c}
\\
\hline
Cluster \#&1&2&3&4&5&6&7&8&9&{10}&Total\\
\hline
Dimension&15& 10 &13& 10& 19&  8& 17& 15&  8 &{14}&129\\
No.\ of exceedances&12& 21& 21 & 9 &28 &10& 26& 28& 14& {33}&202\\
\hline
\end{tabular}
\end{table}

The bottom-up procedure of~\citet{chi2017} runs into the problem that faces of a moderate dimension, say 20, require passing through more than $10^6$ sub-groups.
Nevertheless, their testing criteria can be used to adjust the above discovered faces. 
Instead of the bottom-up approach we can start with a face discovered by clustering and 
apply a greedy strategy to prune or expand this face.
Such greedy pruning is applied to the face~$I_{10}$ and it suggests to exclude stations $2$ and $34$,
since this results in a sharp increase of the associated tail dependence coefficient
from~$0.15$ to~$0.32$.

\section{Graphical models for extremes}
\label{sec:markov}

Statistical modeling of a random vector $X = (X_j: j \in V)$ with moderately large
dimension $|V| = d$ quickly becomes prohibitive because of the complexity of possible dependence
structures between the variables. This is all the more true for extremes where current
parametric models in higher dimension are either simplistic, e.g., the logistic model with just one parameter, or over-parameterized, e.g., the H\"usler--Reiss model with $(d-1)d/2$ parameters; see Section~\ref{sec:classical}. 
Apart from the number of parameters, statistical inference for extreme value models is
challenging even in moderate dimensions.

Conditional independence and graphical models are classical tools for factorizations of high-dimensional
densities, thereby leading to a number of low-dimensional models. 
Such simplified probabilistic structures
facilitate inference and allow the construction of parsimonious parametric models with possibly sparse patterns \citep{Lauritzen, wainwright2008graphical}.
A graphical model for the distribution of $X$ is a set of conditional independence constraints that are encoded by a graph $G = (V,E)$ with vertex set $V$ and edge set $E\subset V \times V$. For disjoint subsets $A,B,C \subset V$, conditional independence between $X_A$ and $X_C$ given $X_B$, denoted by $X_A \ci X_C \mid X_B$, is present if the paths between vertices in $A$ and $C$ are blocked by $B$ in $G$ in a suitable way; we refer to \citet[][Chapter 2]{Lauritzen} and \citet{drt2017} for basic notions of directed and undirected graphs.

A sparse graph with few edges will induce many conditional independencies and the corresponding distribution can be explained by lower-dimensional objects. It is therefore natural to define sparsity in this framework in the following way.
\begin{itemize}
	\item[(S3)]
          A graphical model is sparse if the number of edges is much smaller than
          the number of all possible edges, that is,
          \begin{align}\label{S3}
           |E| \ll d^2. 
          \end{align}
\end{itemize}
In this section we review the recent literature that connect the two fields of extremes and graphical models.

\subsection{Max-stable distributions}
\label{sec:markov_max}

Conditional independence is the basis for the factorization of multivariate densities. For a max-stable random
vector $Z = (Z_j: j \in V)$ \cite{papastathopoulos2016conditional} showed a surprising
negative result on the possibility of such factorizations.
Suppose that $Z$ possesses a positive continuous density and that for disjoint subsets $A,B,C \subset V$, the conditional independence
$$Z_A \ci Z_C \mid Z_B,$$
holds. Then this already implies the unconditional independence $Z_A \ci Z_C$.
This means that no non-trivial conditional independencies are possible in this model
class. 

\begin{figure}
\centering
\includegraphics[clip,height=2.8cm, page=1]{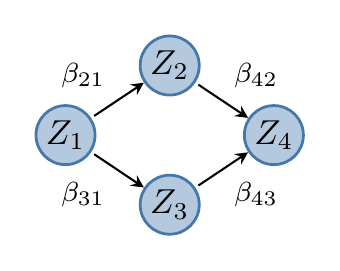}
\caption{Directed acyclic graph $G$ with $d=4$ vertices.}
\label{fig:max-linear}
\end{figure}

While the above result prevents the definition of sparse models for max-stable densities, it does not affect
max-stable models that do not possess densities. An important class of such distributions are the max-linear models defined in Example~\ref{max_linear}. \cite{gis2018} introduce and study a particular sub-class of
max-linear models that are defined on a directed acyclic graph (DAG). A DAG is a graph $G=(V,E)$
with directed edges in $E$ without directed cycles; 
see Figure~\ref{fig:max-linear} for an example in dimension $d=4$. Following the approach of structural equation modeling \citep{spi2000,pea2009}, \cite{gis2018} define a recursive max-linear model on the DAG $G$ by
\begin{align}\label{max_DAG}
  Z_i = \bigvee_{j \in \pa(i)} \beta_{ij} Z_j \vee \beta_{ii} \eps_i, \quad i \in V,
\end{align}
where $\beta_{ij}>0$, $\eps_i$ are independent noise variables with standard Fr\'echet distribution and $\pa(i)$ denotes the graphical parents of the vertex $i$. It is readily verified that the model in Equation~\ref{max_DAG} can be written as a $d$-dimensional max-linear model with $d$ factors defined in Equation~\ref{ML_model} with coefficients $a_{ij}$ that can be derived from the coefficients $\beta_{ij}$.
This also implies that these models have discrete spectral measure with exactly $d$ point masses.

With this construction, the $d$-dimensional max-stable random vector $Z$ in Equation~\ref{max_DAG} does indeed satisfy certain conditional independence relations that are implied by the DAG through so-called $d$-separations. For instance, for the recursive max-linear model in Figure~\ref{fig:max-linear}, we have that $Z_1 \ci Z_4 \mid \{Z_2,Z_3\}$ but we do not have $Z_2 \ci Z_3 \mid \{Z_1,Z_4\}$ \citep{klu2019}.

For a fixed DAG, the estimation of the coefficients $\beta_{ij}$ based on data from $Z$ is challenging due to the discrete nature of max-linear distributions that prohibits standard maximum likelihood methods.
Instead, \cite{yue2014}, \cite{ein2016} and~\cite{ein2018} use M-estimators to circumvent this issue,
and \cite{jan2019} apply the spherical $k$-means clustering described in Section~\ref{sec:clustering} to do inference for max-linear models. 
Alternatively, \cite{gis2019} use a generalized maximum likelihood estimator \citep[see][]{kie1956}. 
The authors of this work also propose a way of learning the graph structure from data.
Further works in this field are \cite{ein2018} who use recursive max-linear model to study European stock market, \cite{buc2020} who consider the model in Equation~\ref{max_DAG} with additional observation errors, and \cite{klu2020} who extend recursive max-linear models to infinite graphs and study connections to percolation theory.

One perspective on recursive max-linear models on DAGs is in terms of Bayesian networks and causality. In this model, large errors propagate through the directed graph deterministically.
In non-extreme statistics, linear structural equation models are common tools in causal inference. For standard Fr\'echet noise variables, similarly to Equation~\ref{max_DAG} we define
\begin{align}\label{sum_DAG}
  X_i = \sum_{j \in \pa(i)} \beta_{ij} X_j + \beta_{ii} \eps_i, \quad i \in V.
\end{align}
Despite their similar structure, the extremal behavior of the linear and max-linear structural equation models in Equations~\ref{sum_DAG} and~\ref{max_DAG}, respectively, may be different, as it is the case for the DAG in Figure~\ref{fig:max-linear} for instance. Such discrepancies disappear after these models are expanded into non-recursive forms.
The causal structure of the model in Equation~\ref{sum_DAG} has been studied in~\citet{gne2019} who propose a greedy search for learning its structure, and define a notion of causal effects in extremes.  

The field of causality for extreme events has become a topic of high interest \citep[e.g.,][]{mha2019,gne2019}, in particular in connection with the attribution of weather extremes in climate science \citep[eg.,][]{han2016,nav2018, nav2020}.

\subsection{Multivariate Pareto distributions}
\label{sec:GM}

\subsubsection{Conditional independence and extremal graphical models}

The negative result by \cite{papastathopoulos2016conditional} presented in Section~\ref{sec:markov_max} does not apply to multivariate Pareto distributions, which are the limits of
threshold exceedances.

Recall from Section~\ref{sec:mrv} the definition of a multivariate Pareto vector $Y = (Y_j: j\in V)$ with support in
$\mathcal L = \{x \in \mathbb R_+^d: \|x\|_\infty \geq 1\}$. Since this space is not a product space,
the classical notions of independence and conditional independence are not applicable.
\cite{eng2018} propose an alternative notion of extremal conditional independence
for a multivariate Pareto distribution $Y$. For any $m\in V$, introduce the auxiliary random vector
$Y^m$ as $Y$ conditioned on the event that $\{Y_m > 1\}$, which has support in the product
space $\{x \in \mathbb R_+^d: x_m \geq  1\}$.
For a partition $\{A,B,C\}$ of $V$, we say that $Y_A$ is conditionally independent of $Y_C$ given $Y_B$ if
\begin{align}\label{Y_CI}
  \forall m\in  \{1,\ldots,d\}: \quad Y^m_A\ci  Y^m_C \mid  Y^m_B.
\end{align}
In this case we write $Y_A\perpe Y_C\mid  Y_B$, where the subscript $\perpe$ indicates extremal independence. When the set $B$ is empty it can be shown that $Y_A\perpe Y_C$ is equivalent to the classical definition of asymptotic extremal independence between $Y_A$ and $Y_C$ \citep{stro2020} as defined in Section~\ref{sec:coef}. The conditional independence notion $\perpe$ is therefore a natural extension to the case more complex conditional extremal independence structures.

From now on we consider undirected graphs $G=(V,E)$.
An extremal graphical model is defined as a multivariate Pareto
distribution $Y$ that satisfies the pairwise Markov property on $G$ with respect to the conditional independence
relation $\perpe$, that is, 
\begin{align}\label{egm}
  Y_i \perpe Y_j \mid Y_{V\setminus\{i,j\}} \quad \text{if } (i,j)\notin E.
\end{align}
Assume now that the graph $G$ is decomposable; see \citet[][Chapter 2]{Lauritzen} for the definition and Figure~\ref{graphical_models} for two examples of decomposable graphs.
Then the analogue of the Hammersley--Clifford theorem holds for the extremal graphical model $Y$ with a positive and continuous density $f_Y$ on $\mathcal L$; see~\cite{eng2018}.
More precisely, the pairwise Markov property in Equation~\ref{egm} on $G$ is equivalent to the factorization
\begin{align}\label{HC_pareto}
  f_{Y}(y)= \frac{1}{\chi_V} {\prod_{C\in \mathcal C} \lambda_C(y_C) \over \prod_{D\in \mathcal D} \lambda_D(y_D)},\quad y\in \mathcal L,
\end{align}
where $\mathcal C$ and $\mathcal D$ are the sets of cliques and intersections between these cliques, respectively.
The factors $\lambda_I$ are the exponent measure densities corresponding to
the vectors $Y_I$ (see Section~\ref{sec:mrv}), and $\chi_V$ is the tail dependence coefficient from Section~\ref{sec:coef}. Moreover, in this case the graph $G$ is necessarily connected, which
means that all components are asymptotically dependent.

It is worthwhile to review some classical statistical models proposed in the literature regarding their sparsity properties.
In fact, many  existing models do not have any conditional independencies and their underlying graphs are  fully connected. This holds for instance for the multivariate logistic distribution, the Dirichlet mixture model \citep{BoldiDavison}, and the pairwise beta distribution \citep{CooleyDavisNaveau}. This observation explains why such models tend to be either too simple or over-parameterized in higher dimensions.
For the simple example where the graph $G$ is a chain, that is, 
\begin{align}\label{chain}
  E = \{\{1,2\},\{2,3\},  \dots, \{d-1,d\}\},
\end{align}
\cite{coles1991modelling} propose a model that factorizes with respect to this graph where all bivariate marginals are logistic, and \cite{smi1997} extend this to general bivariate marginals.
More generally, this relates to the study of extremes of stationary Markov chains where the limiting objects are called tail chains. The multivariate Pareto distributions associated to tail chains factorize with respect to the chain graph; see \cite{smi1992}, \cite{bas2009} and \cite{jan2014}.

\subsubsection{Trees}
\label{sec:trees}
\begin{figure}
  \centering  
  \begin{subfigure}[b]{0.3\textwidth}
  \includegraphics[clip,height=2.8cm, page=2]{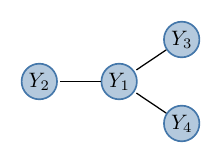}
  \end{subfigure}
    \begin{subfigure}[b]{0.3\textwidth}
  \includegraphics[clip,height=2.8cm, page=3]{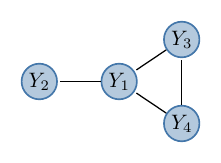}
  \end{subfigure}
    \begin{subfigure}[b]{0.3\textwidth}
  \includegraphics[clip,height=2.8cm, page=4]{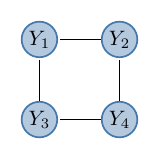}
  \end{subfigure}
  \caption{Three graphical models with $d=4$ vertices. The graphs on the left and in the center are decomposable, whereas the graph on the right is not decomposable.}
  \label{graphical_models}
\end{figure}

A tree $T = (V,E)$ is a graph that is connected and has no cycles; see left-hand side of Figure~\ref{graphical_models} for an example. The number of
edges in a tree equals to $d-1$, all cliques consist of two vertices and the separator sets are singletons. A tree is thus the sparsest model among connected graphs in the sense of the sparsity notion (S3).

An extremal tree model is a multivariate Pareto distribution $Y$ that satisfies the pairwise Markov property in Equation~\ref{egm} on a tree $T$. Such models also appear as the limits of regular varying Markov trees \citep{seg2019}. For extremal tree models, Equation~\ref{HC_pareto} simplifies to
\begin{align}\label{tree_fact}
  f_{Y}(y)=  \frac{1}{\chi_V} \prod_{\{i,j\}\in E} {\lambda_{ij}(y_i,y_j) \over y_i^{-2} y_j^{-2}} \prod_{i\in V} y_i^{-2}.
\end{align}
Apart from characterizing the density of extremal tree models, this formula also
provides a way of constructing new models.
If the tree structure is given or known, for instance through domain knowledge, Equation~\ref{tree_fact} can be used as a recipe to construct sparse, high-dimensional Pareto distributions from bivariate models.  In fact, for any combination of the $d-1$ bivariate
exponent measure densities $(\lambda_{ij}: \{i,j\} \in E)$, Equation~\ref{tree_fact} defines
a valid $d$-dimensional Pareto distribution.
\cite{eng2018} use the density of the bivariate \HR{} distribution (see Example~\ref{ex:HR}) with parameters $\Gamma_{ij}$ for $\lambda_{ij}$ and show that the resulting multivariate density $f_Y$ is again a \HR{} distribution with parameter matrix $\Gamma = (\Gamma_{kl})_{k,l \in V}$ induced by the conditional independence structure, i.e.,
\begin{align}\label{tree_Gamma}
  \Gamma_{kl} =  \sum_{\{i,j\}\in \ph(k,l)} \Gamma_{ij}, \quad k,l \in V,
\end{align}
where $\ph(k,l)$ denotes the set of edges on the unique path from vertex $k$ to vertex~$l$ on the tree $T$.
The number of free parameters in this model is thus $d-1$ and much smaller than the $d(d-1)/2$ parameters in the unrestricted parameter matrix $\Gamma$, thus satisfying the sparsity notion (S3) in Equation~\ref{S3}. This model was also used in \citet{ase2020} in the case where some vertices in the tree are unobserved.

For more flexible statistical modeling it is possible to use different parametric families for the $\lambda_{ij}$, $ \{i,j\} \in E$, or even model them with non-parametric methods. A natural extension of trees are so-called block graphs, i.e., decomposable graphs with singleton separator sets; see, for instance, the graphs on the left-hand side and in the center of Figure~\ref{graphical_models}. For this class of graphical structures, similar formulas as in Equations~\ref{tree_fact} and~\ref{tree_Gamma} hold and the same modular modeling strategy can be used \citep[see][Section 5]{eng2018}.

In most applications, the underlying tree structure is not known and domain knowledge may be unavailable or insufficient. In such cases, the conditional independence structure must be learned from data. A common tool to this end is the notion of a minimum spanning tree.
For each possible edge $\{i,j\}$ between two vertices $i,j \in V$, let $w_{ij}>0$ be a weight, which can be seen as the length of this edge. It is assumed that $w_{ij} = w_{ji}$ and $w_{ii}=0$, $i,j\in V$. The minimum spanning tree is the tree
that minimizes the sum of its edge weights
\begin{align}\label{Tmin}
  T_{\MST} = \argmin_{ T = (V,E)} \sum_{\{i,j\}\in E} w_{ij}. 
\end{align}
The set of all possible trees is very large, but there exist efficient greedy algorithms to solve this problem even for large dimensions $d$ \citep{kruskal1956shortest, pri1957}. The main difficulty is the suitable choice of weights that guarantees that the minimum spanning tree $T_{\MST}$ coincides with the underlying conditional independence tree $T$.

In the non-extremal case of multivariate Gaussian distributions with correlation matrix $(\rho_{ij})_{i,j\in V}$, it can be shown that choosing
$w_{ij}  = - \log |\rho_{ij}|$ (or any monotonically increasing transformation of it) yields the true Gaussian tree structure as the minimum spanning tree \citep[see][]{drt2017}. The assumption of Gaussianity is crucial and the result no longer holds outside of this specific parametric class.

For a multivariate Pareto distribution $Y$ factorizing on a tree $T$, \cite{eng2020} show that the bivariate extremal correlation coefficients introduced in Equation~\ref{eq:chi} can be used for structure learning. In fact, letting 
\begin{align}\label{tree_chi}
  w_{ij} =  - \log \chi_{ij}, \quad i,j \in V,
\end{align}
the minimum spanning tree in Equation~\ref{Tmin} satisfies $T_{\MST} = T$. It should be noted that this result holds regardless of the distribution of $Y$ and no assumption on a specific parametric model class is required. This is quite surprising, since it is stronger than in the classical, non-extremal theory of trees. \cite{eng2020} further introduce a new summary statistic, the extremal variogram, which can also be used in Equation~\ref{Tmin} to consistently recover the extremal tree structure, and which tends to be more accurate in finite samples. An alternative approach is to use likelihood based methods to learn the tree structure. This requires to specify a parametric model class, but it also allows to learn certain block graph structures by a forward selection algorithm \cite[see][]{eng2018}. 

\subsubsection{\HR{} graphical models}

Going beyond trees and block graphs requires a distributional assumption on the multivariate Pareto distribution. The \HR{} distribution, introduced in Example~\ref{ex:HR}, can be seen as the natural analogue of Gaussian distributions in the world of asymptotically dependent extremes.

While for a multivariate Gaussian distribution with covariance matrix $\Sigma$ the conditional independence structure can be identified from the zeros on the precision matrix $\Sigma^{-1}$, for
\HR{} distributions the conditionally negative definite parameter matrix $\Gamma$ plays the key role.
Recall from Equation~\ref{sigma_k} the matrix $\Sigma^{(m)}$, which is of dimension $(d-1)\times (d-1)$ since the $m$th row and column are omitted. \citet[][Proposition 3]{eng2018} show that for any $m \in V$, the inverse $K^{(m)}$ of the matrix $\Sigma^{(m)}$ satisfies
  \begin{align}\label{CI_HR}
    Y_i\perpe Y_j\mid  Y_{V \setminus \{i,j\}} \quad \iff \quad
    \begin{cases}
       K^{(m)}_{ij}= 0,  &\text{ if } i,j \neq m,\\
       \sum_{l\neq m} K^{(m)}_{il} = 0, & \text{ if } i \neq m, j = m. 
    \end{cases} 
  \end{align}
  For any $m \in V$, the single matrix $K^{(m)}$ contains all information on the extremal graphical structure. Edges between vertices not including the $m$th vertex correspond to zeros on the off-diagonal, while edges including the $m$th vertex correspond to zero row sums. This result holds for any graph, even if it is not decomposable.

\HR{} distributions are the finite dimensional distributions of Brown--Resnick processes, a widely used model for spatial extreme events parameterized by variogram functions \citep{bro1977, kab2009}. Equation~\ref{CI_HR} can be used to see that most popular parametric classes of variogram functions yield models that do not have any conditional independencies. An exception is the original Brown--Resnick process introduced in \cite{bro1977} whose finite dimensional distributions factorize on the chain graph in Equation~\ref{chain}.

\subsection{Application to flood risk}\label{graph_application}

In Section~\ref{conc_application} the group of stations $I_{10}$ was identified as the one manifesting most frequent concomitant extremes; it was then adjusted by removing stations $2$ and $34$ as suggested in Section~\ref{conc_application}. For illustration purposes, we add the larger downstream stations $63$, $64$ and $67$, and then analyze this group of $15$ stations with the extremal graphical models in Section~\ref{sec:GM}. We use the
R \citep{R2019} implementation of the package \texttt{graphicalExtremes} \citep{graphicalExtremes} for structure learning and model fitting.

We do not use any geographical information on the locations but learn the graph structure from the data. First, we estimate the minimum spanning tree $T_{\MST}$ with weights $-\log\widehat \chi_{ij}$ based on the empirical tail dependence coefficients; see Section~\ref{sec:trees}. We choose to use a \HR{} distribution on the tree structure, and then extend this model by adding edges in a greedy way while staying in the class of block graphs with cliques of maximal size three \citep[see][Section 6]{eng2018}. Fitting these models requires maximization of bivariate and trivariate \HR{} densities with censoring for non-extreme components \citep[see][]{led1997}. The greedy forward selection is based on the AIC score. 

\begin{figure}
  \centering
    \begin{subfigure}[b]{0.49\textwidth}
  \includegraphics[width=\textwidth]{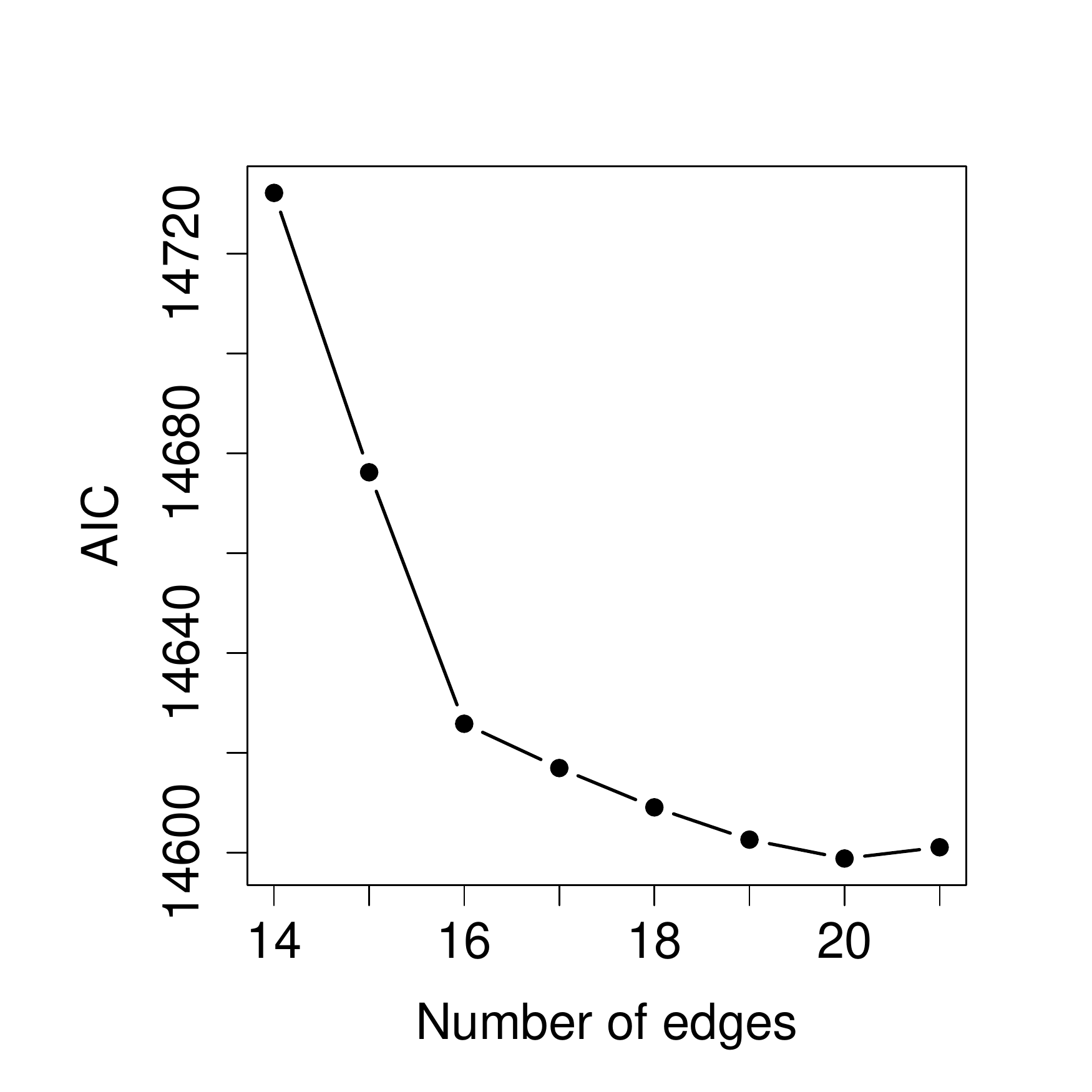}
  \end{subfigure}
      \begin{subfigure}[b]{0.49\textwidth}
  \includegraphics[width=\textwidth]{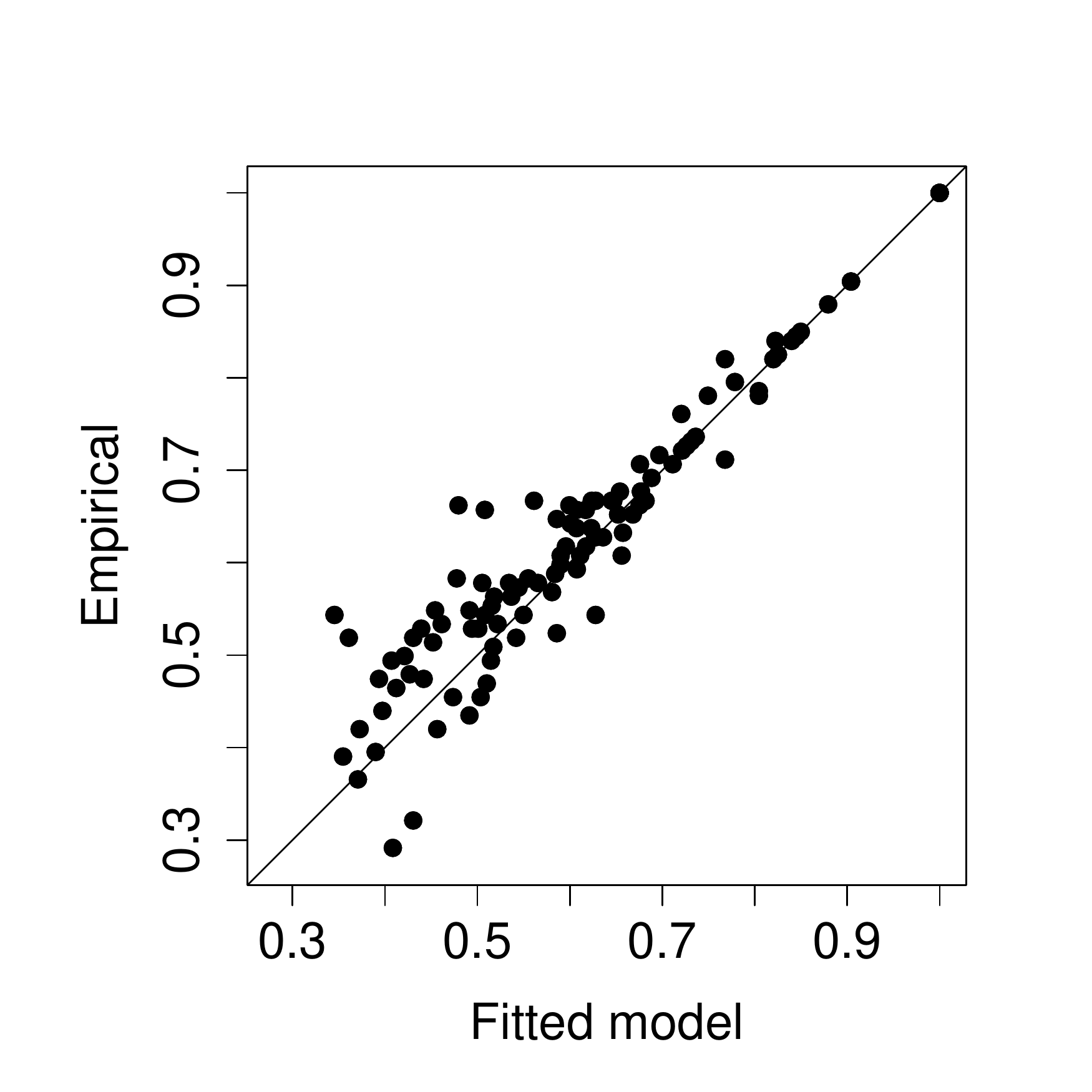}
  \end{subfigure}
  \caption{Left: AIC values for extremal graphical models for the river data set with an increasing
number of edges, starting from the minimum spanning tree. Right: empirically estimated $\widehat \chi_{ij}$ coefficients for all pairs of stations against those implied by the fitted graphical model minimizing the AIC.}
  \label{model_fit}
\end{figure}

\begin{figure}
  \centering
  \includegraphics[clip, width=0.7\textwidth, page=5]{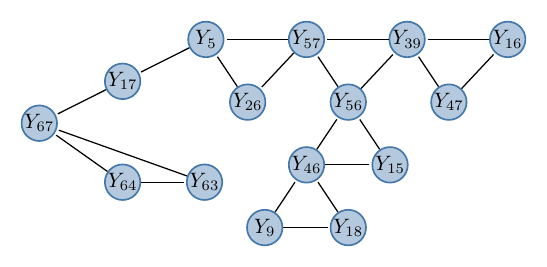}
  \caption{The estimated block graph corresponding to the model with the optimal AIC.}
  \label{model_graph}
\end{figure}

The AIC curve of the resulting model fits is shown on the left-hand side of Figure~\ref{model_fit}. The best model minimizing the AIC has $20$ edges and is significantly better than the simpler tree model.
Note that the number of 20 free parameters in this block graph model is much lower than the $105$ parameters in a dense \HR{} model with $d=15$. Moreover, because of the density decomposition in Equation~\ref{HC_pareto}, the parameters on each of the cliques can be estimated separately, rendering inference much more efficient \citep[see][Section 5]{eng2018}. The estimated graph structure of the best model is shown in Figure~\ref{model_graph}. We note that the graph does only roughly reflect the flow connections in the river network in Figure~\ref{fig:basin}. This is not a contradiction, since other effects such as spatial precipitation events may affect dependence between the peak flows. For instance, the fact that the stations $63$ and $64$ are not directly connected to station $47$ may be due to the large lakes that dampen the largest discharges.
The right-hand side of Figure~\ref{model_fit} compares empirical estimates $\widehat \chi_{ij}$ of the tail dependence coefficients between all $15$ stations with those implied by the best fitted model.
The plot underlines that the sparse graphical \HR{} model captures well the extremal dependence in the data.

\begin{issues}[FUTURE ISSUES]
\begin{enumerate}
\item Adaptation of further methods from multivariate analysis and machine learning for rare event modeling.
\item Automatic and efficient search algorithms to select out of all $2^d-1$ possible faces those with $\Lambda$ mass.
\item Structure estimation methods for extremal graphical models beyond trees and automatic selection of the degree of sparsity.
\item Modeling sparse structures in sub-asymptotic extremes for asymptotically independent data.
\item Flexible sparse models for mixtures of asymptotic independence and dependence.
\item Modeling and detection of causal effects in the distributional tails.  
\item Methods for extreme value analysis in high-dimensional settings $d \gg n$.   
\end{enumerate}
\end{issues}

\section*{ACKNOWLEDGMENTS}
J.\ Ivanovs gratefully acknowledges financial support of Sapere Aude Starting Grant 8049-00021B ``Distributional Robustness in Assessment of Extreme Risk''.

\bibliographystyle{ar-style1}
\bibliography{references}

\begin{thebibliography}{}
\expandafter\ifx\csname natexlab\endcsname\relax\def\natexlab#1{#1}\fi

\bibitem[Aas et~al.(2009)Aas, Czado, Frigessi \& Bakken]{aas2009}
Aas K, Czado C, Frigessi A, Bakken H. 2009.
Pair-copula constructions of multiple dependence.
\textit{Insurance: Mathematics and Economics} 44:182 -- 198

\bibitem[Agrawal et~al.(1994)Agrawal, Srikant et~al.]{agr1994}
Agrawal R, Srikant R, et~al. 1994.
Fast algorithms for mining association rules, In \textit{Proceedings of the
  20th International Conference on Very Large Data Bases, VLDB}, vol. 1215,\
  pp.  487--499

\bibitem[Anderson(2003)]{anderson_book}
Anderson TW. 2003.
An introduction to multivariate statistical analysis.
Wiley Series in Probability and Statistics. Wiley-Interscience [John Wiley \&
  Sons], Hoboken, NJ, 3rd ed.

\bibitem[Asadi et~al.(2015)Asadi, Davison \& Engelke]{asadi2015extremes}
Asadi P, Davison AC, Engelke S. 2015.
Extremes on river networks.
\textit{The Annals of Applied Statistics} 9:2023--2050

\bibitem[Asadi et~al.(2018)Asadi, Engelke \& Davison]{asa2018}
Asadi P, Engelke S, Davison A. 2018.
Optimal regionalization of extreme value distributions for flood estimation.
\textit{Journal of Hydrology} 556:182--193

\bibitem[Asenova et~al.(2020)Asenova, Mazo \& Segers]{ase2020}
Asenova S, Mazo G, Segers J. 2020.
Inference on extremal dependence in a latent {M}arkov tree model attracted to a
  {H}{\"u}sler--{R}eiss distribution.
Available from \texttt{https://arxiv.org/abs/2001.09510}

\bibitem[Balkema \& de~Haan(1974)]{bal74}
Balkema AA, de~Haan L. 1974.
Residual life time at great age.
\textit{Ann. Probability} 2:792--804

\bibitem[Basrak et~al.(2002)Basrak, Davis \&
  Mikosch]{basrak2002characterization}
Basrak B, Davis RA, Mikosch T. 2002.
A characterization of multivariate regular variation.
\textit{The Annals of Applied Probability} 12:908--920

\bibitem[Basrak \& Segers(2009)]{bas2009}
Basrak B, Segers J. 2009.
Regularly varying multivariate time series.
\textit{Stochastic Processes and their Applications} 119:1055 -- 1080

\bibitem[Beirlant et~al.(2004)Beirlant, Goegebeur, Teugels \& Segers]{ber2004}
Beirlant J, Goegebeur Y, Teugels J, Segers J. 2004.
Statistics of {E}xtremes.
Wiley Series in Probability and Statistics. John Wiley \& Sons, Ltd.,
  Chichester

\bibitem[Bernard et~al.(2013)Bernard, Naveau, Vrac \&
  Mestre]{bernard2013clustering}
Bernard E, Naveau P, Vrac M, Mestre O. 2013.
Clustering of maxima: Spatial dependencies among heavy rainfall in {F}rance.
\textit{Journal of Climate} 26:7929--7937

\bibitem[Blanchard et~al.(2007)Blanchard, Bousquet \&
  Zwald]{blanchard2007statistical}
Blanchard G, Bousquet O, Zwald L. 2007.
Statistical properties of kernel principal component analysis.
\textit{Machine Learning} 66:259--294

\bibitem[Boldi \& Davison(2007)]{BoldiDavison}
Boldi MO, Davison AC. 2007.
A mixture model for multivariate extremes.
\textit{Journal of the Royal Statistical Society. Series B. Statistical
  Methodology} 69:217--229

\bibitem[Brown \& Resnick(1977)]{bro1977}
Brown BM, Resnick SI. 1977.
Extreme values of independent stochastic processes.
\textit{J. Appl. Probab.} 14:732--739

\bibitem[B{\"u}cher et~al.(2014)B{\"u}cher, Segers \& Volgushev]{BSV2014}
B{\"u}cher A, Segers J, Volgushev S. 2014.
When uniform weak convergence fails: Empirical processes for dependence
  functions and residuals via epi- and hypographs.
\textit{Ann. Stat.} 42:1598--1634

\bibitem[Buck \& Kl\"{u}ppelberg(2020)]{buc2020}
Buck J, Kl\"{u}ppelberg C. 2020.
Recursive max-linear models with propagating noise.
Available from \texttt{https://arxiv.org/abs/2003.00362}.

\bibitem[Chautru(2015)]{cha2015}
Chautru E. 2015.
Dimension reduction in multivariate extreme value analysis.
\textit{Electron. J. Statist.} 9:383--418

\bibitem[Chiapino \& Sabourin(2017)]{chi2017}
Chiapino M, Sabourin A. 2017.
Feature clustering for extreme events analysis, with application to extreme
  stream-flow data, In \textit{New Frontiers in Mining Complex Patterns}, eds.
  A~Appice, M~Ceci, C~Loglisci, E~Masciari, ZW~Ra{\'{s}},\ pp.  132--147, Cham:
  Springer International Publishing

\bibitem[Chiapino et~al.(2019)Chiapino, Sabourin \& Segers]{chi2019}
Chiapino M, Sabourin A, Segers J. 2019.
Identifying groups of variables with the potential of being large
  simultaneously.
\textit{Extremes} 22:193--222

\bibitem[Coles et~al.(1999)Coles, Heffernan \& Tawn]{col1999}
Coles S, Heffernan J, Tawn J. 1999.
Dependence measures for extreme value analyses.
\textit{Extremes} 2:339--365

\bibitem[Coles(2001)]{coles2001introduction}
Coles SG. 2001.
An introduction to statistical modeling of extreme values.
Springer Series in Statistics. Springer

\bibitem[Coles \& Tawn(1991)]{coles1991modelling}
Coles SG, Tawn JA. 1991.
Modelling extreme multivariate events.
\textit{Journal of the Royal Statistical Society. Series B. Methodological}
  53:377--392

\bibitem[Cooley et~al.(2010)Cooley, Davis \& Naveau]{CooleyDavisNaveau}
Cooley D, Davis RA, Naveau P. 2010.
The pairwise beta distribution: A flexible parametric multivariate model for
  extremes.
\textit{Journal of Multivariate Analysis} 101:2103--2117

\bibitem[Cooley \& Thibaud(2019)]{coo2019}
Cooley D, Thibaud E. 2019.
{Decompositions of dependence for high-dimensional extremes}.
\textit{Biometrika} 106:587--604

\bibitem[Davis et~al.(2013)Davis, Kl{\"u}ppelberg \& Steinkohl]{dav2013}
Davis RA, Kl{\"u}ppelberg C, Steinkohl C. 2013.
Statistical inference for max-stable processes in space and time.
\textit{J. R. Stat. Soc. Ser. B Stat. Methodol.} 75:791--819

\bibitem[Davison \& Huser(2015)]{dav2015}
Davison A, Huser R. 2015.
Statistics of extremes.
\textit{Annual Review of Statistics and Its Application} 2:203--235

\bibitem[Davison et~al.(2012)Davison, Padoan \& Ribatet]{dav2012b}
Davison AC, Padoan SA, Ribatet M. 2012.
Statistical modeling of spatial extremes.
\textit{Statist. Sci.} 27:161--186

\bibitem[de~Fondeville \& Davison(2018)]{def2018}
de~Fondeville R, Davison AC. 2018.
High-dimensional peaks-over-threshold inference.
\textit{Biometrika} 105:575--592

\bibitem[de~Haan \& Ferreira(2006)]{deh2006a}
de~Haan L, Ferreira A. 2006.
Extreme value theory.
New York: Springer

\bibitem[de~Haan \& Zhou(2011)]{deHaa11}
de~Haan L, Zhou C. 2011.
Extreme residual dependence for random vectors and processes.
\textit{Adv. in Appl. Probab.} 43:217--242

\bibitem[Dhillon \& Modha(2001)]{dhi2001}
Dhillon IS, Modha DS. 2001.
Concept decompositions for large sparse text data using clustering.
\textit{Machine Learning} 42:143--175

\bibitem[Dieker \& Mikosch(2015)]{die2015}
Dieker AB, Mikosch T. 2015.
Exact simulation of {B}rown--{R}esnick random fields at a finite number of
  locations.
\textit{Extremes} 18:301--314

\bibitem[Dombry et~al.(2016)Dombry, Engelke \& Oesting]{dom2016}
Dombry C, Engelke S, Oesting M. 2016.
Exact simulation of max-stable processes.
\textit{Biometrika} 103:303--317

\bibitem[Dombry et~al.(2017{\natexlab{a}})Dombry, Engelke \& Oesting]{dom2016b}
Dombry C, Engelke S, Oesting M. 2017{\natexlab{a}}.
Asymptotic properties of the maximum likelihood estimator for multivariate
  extreme value distributions.
Available from \texttt{https://arxiv.org/abs/1612.05178}

\bibitem[Dombry et~al.(2017{\natexlab{b}})Dombry, Engelke \& Oesting]{dom2016a}
Dombry C, Engelke S, Oesting M. 2017{\natexlab{b}}.
Bayesian inference for multivariate extreme value distributions.
\textit{Electronic Journal of Statistics} 11:4813--4844

\bibitem[Dombry et~al.(2013)Dombry, Eyi-Minko \& Ribatet]{dom2013}
Dombry C, Eyi-Minko F, Ribatet M. 2013.
Conditional simulation of max-stable processes.
\textit{Biometrika} 100:111--124

\bibitem[Drees \& Huang(1998)]{DH1998}
Drees H, Huang X. 1998.
Best attainable rates of convergence for estimators of the stable tail
  dependence function.
\textit{Journal of Multivariate Analysis} 64:25--46

\bibitem[Drees \& Sabourin(2019)]{dre2019}
Drees H, Sabourin A. 2019.
Principal component analysis for multivariate extremes.
Available from \texttt{https://arxiv.org/abs/1906.11043}.

\bibitem[Drton \& Maathuis(2017)]{drt2017}
Drton M, Maathuis MH. 2017.
Structure learning in graphical modeling.
\textit{Annual Review of Statistics and Its Application} 4:365--393

\bibitem[Duchi et~al.(2008)Duchi, Shalev-Shwartz, Singer \&
  Chandra]{duchi2008efficient}
Duchi J, Shalev-Shwartz S, Singer Y, Chandra T. 2008.
Efficient projections onto the $l_1$-ball for learning in high dimensions, In
  \textit{Proceedings of the 25th international conference on Machine
  Learning},\ pp.  272--279, ACM

\bibitem[Eastoe \& Tawn(2012)]{Eas12}
Eastoe EF, Tawn JA. 2012.
Modelling the distribution of the cluster maxima of exceedances of
  subasymptotic thresholds.
\textit{Biometrika} 99:43--55

\bibitem[Einmahl et~al.(2012{\natexlab{a}})Einmahl, Krajina \& Segers]{EKS2012}
Einmahl J, Krajina A, Segers J. 2012{\natexlab{a}}.
An m-estimator for tail dependence in arbitrary dimensions.
\textit{Ann. Stat.} 40:1764--1793

\bibitem[Einmahl et~al.(2016)Einmahl, Kiriliouk, Krajina \& Segers]{ein2016}
Einmahl JHJ, Kiriliouk A, Krajina A, Segers J. 2016.
An {$M$}-estimator of spatial tail dependence.
\textit{J. R. Stat. Soc. Ser. B. Stat. Methodol.} 78:275--298

\bibitem[Einmahl et~al.(2018)Einmahl, Kiriliouk \& Segers]{ein2018}
Einmahl JHJ, Kiriliouk A, Segers J. 2018.
A continuous updating weighted least squares estimator of tail dependence in
  high dimensions.
\textit{Extremes} 21:205--233

\bibitem[Einmahl et~al.(2012{\natexlab{b}})Einmahl, Krajina \& Segers]{ein2012}
Einmahl JHJ, Krajina A, Segers J. 2012{\natexlab{b}}.
An {M}-estimator for tail dependence in arbitrary dimensions.
\textit{Ann. Statist.} 40:1764--1793

\bibitem[Embrechts et~al.(1997)Embrechts, Kl\"{u}ppelberg \& Mikosch]{emb1997}
Embrechts P, Kl\"{u}ppelberg C, Mikosch T. 1997.
Modelling extremal events: for insurance and finance.
London: Springer

\bibitem[Engelke et~al.(2019{\natexlab{a}})Engelke, de~Fondeville \&
  Oesting]{eng2017a}
Engelke S, de~Fondeville R, Oesting M. 2019{\natexlab{a}}.
Extremal behaviour of aggregated data with an application to downscaling.
\textit{Biometrika} 106:127--144

\bibitem[Engelke \& Hitz(2019)]{eng2018}
Engelke S, Hitz A. 2019.
Graphical models for extremes (with discussion).
\textit{Accepted in J. R. Stat. Soc. Ser. B Stat. Methodol.} Available from
  \texttt{https://arxiv.org/abs/1812.01734}.

\bibitem[Engelke et~al.(2019{\natexlab{b}})Engelke, Hitz \&
  Gnecco]{graphicalExtremes}
Engelke S, Hitz SA, Gnecco N. 2019{\natexlab{b}}.
graphical{E}xtremes: Statistical methodology for graphical extreme value
  models.
Available from \texttt{https://CRAN.R-project.org/package=graphicalExtremes}, R
  package version 0.1.0

\bibitem[Engelke et~al.(2015)Engelke, Malinowski, Kabluchko \&
  Schlather]{Engelke2015}
Engelke S, Malinowski A, Kabluchko Z, Schlather M. 2015.
Estimation of {H}{\"u}sler--{R}eiss distributions and {B}rown--{R}esnick
  processes.
\textit{Journal of the Royal Statistical Society. Series B. Methodological}
  77:239--265

\bibitem[Engelke et~al.(2019{\natexlab{c}})Engelke, Opitz \&
  Wadsworth]{eng2018a}
Engelke S, Opitz T, Wadsworth J. 2019{\natexlab{c}}.
Extremal dependence of random scale constructions.
\textit{Extremes} 22:623--666

\bibitem[Engelke \& Volgushev(2020)]{eng2020}
Engelke S, Volgushev S. 2020.
The extremal variogram and tree structure learning.
In preparation

\bibitem[Fisher \& Tippett(1928)]{fis1928}
Fisher RA, Tippett LHC. 1928.
Limiting forms of the frequency distribution of the largest or smallest member
  of a sample, In \textit{Mathematical Proceedings of the Cambridge
  Philosophical Society}, vol.~24,\ pp.  180--190, Cambridge University Press

\bibitem[Gissibl \& Kl{\"u}ppelberg(2018)]{gis2018}
Gissibl N, Kl{\"u}ppelberg C. 2018.
Max-linear models on directed acyclic graphs.
\textit{Bernoulli} 24:2693--2720

\bibitem[Gissibl et~al.(2019)Gissibl, Kl{\"u}ppelberg \& Lauritzen]{gis2019}
Gissibl N, Kl{\"u}ppelberg C, Lauritzen S. 2019.
Identifiability and estimation of recursive max-linear models.
Available from \texttt{https://arxiv.org/abs/1901.03556}.

\bibitem[Gnecco et~al.(2019)Gnecco, Meinshausen, Peters \& Engelke]{gne2019}
Gnecco N, Meinshausen N, Peters J, Engelke S. 2019.
Causal discovery in heavy-tailed models.
Available from \texttt{https://arxiv.org/abs/1908.05097}.

\bibitem[Goix et~al.(2016)Goix, Sabourin \& Clémençon]{goi2016}
Goix N, Sabourin A, Clémençon S. 2016.
Sparse representation of multivariate extremes with applications to anomaly
  ranking, In \textit{Proceedings of the 19th International Conference on
  Artificial Intelligence and Statistics (AISTATS)}. JMLR: W\&CP

\bibitem[Goix et~al.(2017)Goix, Sabourin \& Clémençon]{goi2017}
Goix N, Sabourin A, Clémençon S. 2017.
Sparse representation of multivariate extremes with applications to anomaly
  detection.
\textit{Journal of Multivariate Analysis} 161:12 -- 31

\bibitem[Gudendorf \& Segers(2010)]{seg2010}
Gudendorf G, Segers J. 2010.
Extreme-value copulas. In \textit{Copula Theory and Its Applications}.
  Springer,  127--145

\bibitem[Hannart et~al.(2016)Hannart, Pearl, Otto, Naveau \& Ghil]{han2016}
Hannart A, Pearl J, Otto FEL, Naveau P, Ghil M. 2016.
Causal counterfactual theory for the attribution of weather and climate-related
  events.
\textit{Bulletin of the American Meteorological Society} 97:99--110

\bibitem[Heffernan \& Tawn(2004)]{HeffernanTawn2004}
Heffernan JE, Tawn JA. 2004.
A conditional approach for multivariate extreme values (with discussion).
\textit{Journal of the Royal Statistical Society: Series B (Statistical
  Methodology)} 66:497--546

\bibitem[Hill(1975)]{hil1975}
Hill BM. 1975.
A simple general approach to inference about the tail of a distribution.
\textit{Ann. Statist.} 3:1163--1174

\bibitem[Hitz \& Evans(2016)]{Hitz2015}
Hitz SA, Evans JR. 2016.
One-component regular variation and graphical modeling of extremes.
\textit{Journal of Applied Probability} 53:733--746

\bibitem[Huang(1992)]{H1992}
Huang X. 1992.
Statistics of bivariate extreme value theory.
Ph.D. thesis, Erasmus University Rotterdam

\bibitem[Huser et~al.(2019)Huser, Dombry, Ribatet \& Genton]{hus2019}
Huser R, Dombry C, Ribatet M, Genton MG. 2019.
Full likelihood inference for max-stable data.
\textit{Stat} 8:e218

\bibitem[Huser \& Wadsworth(2019)]{HuserWadsworth2017}
Huser R, Wadsworth JL. 2019.
Modeling spatial processes with unknown extremal dependence class.
\textit{J. Amer. Statist. Assoc.} 114:434--444

\bibitem[H{\"u}sler \& Reiss(1989)]{Husler1989}
H{\"u}sler J, Reiss RD. 1989.
Maxima of normal random vectors: between independence and complete dependence.
\textit{Statistics \& Probability Letters} 7:283--286

\bibitem[Janssen \& Segers(2014)]{jan2014}
Janssen A, Segers J. 2014.
{M}arkov tail chains.
\textit{Journal of Applied Probability} 51:1133–1153

\bibitem[Janssen \& Wan(2019)]{jan2019}
Janssen A, Wan P. 2019.
{$k$}-means clustering of extremes.
Available from \texttt{https://arxiv.org/abs/1904.02970}.

\bibitem[Jung et~al.(2012)Jung, Dryden \& Marron]{jung12}
Jung S, Dryden IL, Marron JS. 2012.
Analysis of principal nested spheres.
\textit{Biometrika} 99:551--568

\bibitem[Kabluchko et~al.(2009)Kabluchko, Schlather \& de~Haan]{kab2009}
Kabluchko Z, Schlather M, de~Haan L. 2009.
Stationary max-stable fields associated to negative definite functions.
\textit{Ann. Probab.} 37:2042--2065

\bibitem[Katz et~al.(2002)Katz, Parlange \& Naveau]{kat2002}
Katz RW, Parlange MB, Naveau P. 2002.
Statistics of extremes in hydrology.
\textit{Advances in Water Resources} 25:1287--1304

\bibitem[Keef et~al.(2009)Keef, Tawn \& Svensson]{kee2009}
Keef C, Tawn J, Svensson C. 2009.
Spatial risk assessment for extreme river flows.
\textit{J. R. Stat. Soc. Ser. C. Appl. Stat.} 58:601--618

\bibitem[Kiefer \& Wolfowitz(1956)]{kie1956}
Kiefer J, Wolfowitz J. 1956.
Consistency of the maximum likelihood estimator in the presence of infinitely
  many incidental parameters.
\textit{Ann. Math. Statist.} 27:887--906

\bibitem[Kl\"{u}ppelberg et~al.(2015)Kl\"{u}ppelberg, Haug \&
  Kuhn]{haug2009dimension}
Kl\"{u}ppelberg C, Haug S, Kuhn G. 2015.
Copula structure analysis based on extreme dependence.
\textit{Stat. Interface} 8:93--107

\bibitem[Kl\"{u}ppelberg \& Lauritzen(2019)]{klu2019}
Kl\"{u}ppelberg C, Lauritzen S. 2019.
Bayesian networks for max-linear models.
Available from \texttt{https://arxiv.org/abs/1901.03948}.

\bibitem[Kl{\"u}ppelberg \& S{\"o}nmez(2020)]{klu2020}
Kl{\"u}ppelberg C, S{\"o}nmez E. 2020.
Max-linear models on infinite graphs generated by bernoulli bond percolation.
Available from \texttt{https://arxiv.org/abs/1804.06102}

\bibitem[Kruskal(1956)]{kruskal1956shortest}
Kruskal Jr. JB. 1956.
On the shortest spanning subtree of a graph and the traveling salesman problem.
\textit{Proceedings of the American Mathematical Society} 7:48--50

\bibitem[Larsson \& Resnick(2012)]{lar2012}
Larsson M, Resnick SI. 2012.
Extremal dependence measure and extremogram: the regularly varying case.
\textit{Extremes} 15:231--256

\bibitem[Lauritzen(1996)]{Lauritzen}
Lauritzen SL. 1996.
Graphical models.
Oxford University Press

\bibitem[Ledford \& Tawn(1997)]{led1997}
Ledford AW, Tawn JA. 1997.
Modelling dependence within joint tail regions.
\textit{Journal of the Royal Statistical Society: Series B (Statistical
  Methodology)} 59:475--499

\bibitem[Lee \& Joe(2018)]{lee2018}
Lee D, Joe H. 2018.
Multivariate extreme value copulas with factor and tree dependence structures.
\textit{Extremes} 21:147--176

\bibitem[Lehtomaa \& Resnick(2019)]{lehtomaa2019asymptotic}
Lehtomaa J, Resnick S. 2019.
Asymptotic independence and support detection techniques for heavy-tailed
  multivariate data.
Available from \texttt{https://arxiv.org/abs/1904.00917}.

\bibitem[Lindskog et~al.(2014)Lindskog, Resnick \& Roy]{lindskog14}
Lindskog F, Resnick SI, Roy J. 2014.
Regularly varying measures on metric spaces: hidden regular variation and
  hidden jumps.
\textit{Probab. Surv.} 11:270--314

\bibitem[McNeil et~al.(2015)McNeil, Frey \& Embrechts]{mcn2015}
McNeil AJ, Frey R, Embrechts P. 2015.
Quantitative risk management: Concepts, techniques and tools.
Princeton University Press

\bibitem[Meyer \& Wintenberger(2019)]{mey2019}
Meyer N, Wintenberger O. 2019.
Sparse regular variation.
Available from \texttt{https://arxiv.org/abs/1907.00686}.

\bibitem[Mhalla et~al.(2019)Mhalla, Chavez-Demoulin \& Dupuis]{mha2019}
Mhalla L, Chavez-Demoulin V, Dupuis DJ. 2019.
Causal mechanism of extreme river discharges in the upper {D}anube basin
  network.
Available from \texttt{https://arxiv.org/abs/1907.03555}.

\bibitem[Naveau et~al.(2020)Naveau, Hannart \& Ribes]{nav2020}
Naveau P, Hannart A, Ribes A. 2020.
{Statistical methods for extreme event attribution in climate science}.
\textit{Annual Review of Statistics and Its Application} {To appear}

\bibitem[Naveau et~al.(2018)Naveau, Ribes, Zwiers, Hannart, Tuel \&
  Yiou]{nav2018}
Naveau P, Ribes A, Zwiers F, Hannart A, Tuel A, Yiou P. 2018.
Revising return periods for record events in a climate event attribution
  context.
\textit{Journal of Climate} 31:3411--3422

\bibitem[Opitz(2013)]{opi2013}
Opitz T. 2013.
Extremal {$t$} processes: Elliptical domain of attraction and a spectral
  representation.
\textit{J. Multivariate Anal.} 122:409--413

\bibitem[Padoan et~al.(2010)Padoan, Ribatet \& Sisson]{pad2010}
Padoan SA, Ribatet M, Sisson SA. 2010.
Likelihood-based inference for max-stable processes.
\textit{J. Amer. Statist. Assoc.} 105:263--277

\bibitem[Papastathopoulos \& Strokorb(2016)]{papastathopoulos2016conditional}
Papastathopoulos I, Strokorb K. 2016.
Conditional independence among max-stable laws.
\textit{Statistics \& Probability Letters} 108:9--15

\bibitem[Papastathopoulos et~al.(2017)Papastathopoulos, Strokorb, Tawn \&
  Butler]{pap2017}
Papastathopoulos I, Strokorb K, Tawn JA, Butler A. 2017.
Extreme events of {M}arkov chains.
\textit{Advances in Applied Probability} 49:134--161

\bibitem[Pearl(2009)]{pea2009}
Pearl J. 2009.
Causality.
Cambridge University Press, Cambridge, 2nd ed.
Models, reasoning, and inference

\bibitem[Peng(1999)]{pen99}
Peng L. 1999.
Estimation of the coefficient of tail dependence in bivariate extremes.
\textit{Statist. Probab. Lett.} 43:399--409

\bibitem[Pickands(1975)]{Pickands1975}
Pickands III J. 1975.
Statistical inference using extreme order statistics.
\textit{The Annals of Statistics} 3:119--131

\bibitem[Poon et~al.(2004)Poon, Rockinger \& Tawn]{poo2004}
Poon SH, Rockinger M, Tawn J. 2004.
Extreme value dependence in financial markets: Diagnostics, models, and
  financial implications.
\textit{Rev. Financ. Stud.} 17:581--610

\bibitem[Prim(1957)]{pri1957}
Prim RC. 1957.
Shortest connection networks and some generalizations.
\textit{Bell System Technical Journal} 36:1389--1401

\bibitem[{R Core Team}(2019)]{R2019}
{R Core Team}. 2019.
R: A language and environment for statistical computing.
R Foundation for Statistical Computing, Vienna, Austria

\bibitem[Ramos \& Ledford(2009)]{ram09}
Ramos A, Ledford A. 2009.
A new class of models for bivariate joint tails.
\textit{J. R. Stat. Soc. Ser. B Stat. Methodol.} 71:219--241

\bibitem[Reich \& Shaby(2012)]{rei2012}
Reich BJ, Shaby BA. 2012.
A hierarchical max-stable spatial model for extreme precipitation.
\textit{Ann. Appl. Stat.} 6:1430--1451

\bibitem[Resnick(2008)]{res2008}
Resnick SI. 2008.
Extreme values, regular variation and point processes.
New York: Springer

\bibitem[Rootz{\'e}n \& Tajvidi(2006)]{roo2006}
Rootz{\'e}n H, Tajvidi N. 2006.
Multivariate generalized {P}areto distributions.
\textit{Bernoulli} 12:917--930

\bibitem[Samorodnitsky et~al.(2016)Samorodnitsky, Resnick, Towsley, Davis,
  Willis \& Wan]{sam2016}
Samorodnitsky G, Resnick S, Towsley D, Davis R, Willis A, Wan P. 2016.
Nonstandard regular variation of in-degree and out-degree in the preferential
  attachment model.
\textit{Journal of Applied Probability} 53:146–161

\bibitem[Saunders et~al.(2019)Saunders, Stephenson \& Karoly]{Saunders2019}
Saunders KR, Stephenson AG, Karoly DJ. 2019.
A regionalisation approach for rainfall based on extremal dependence.
Available from \texttt{https://arxiv.org/abs/1907.05750}.

\bibitem[Schlather(2002)]{sch2002}
Schlather M. 2002.
Models for stationary max-stable random fields.
\textit{Extremes} 5:33--44

\bibitem[Schlather \& Tawn(2002)]{schlather02}
Schlather M, Tawn J. 2002.
Inequalities for the extremal coefficients of multivariate extreme value
  distributions.
\textit{Extremes} 5:87--102

\bibitem[Seber(1984)]{seber_book}
Seber GAF. 1984.
Multivariate observations.
Wiley Series in Probability and Mathematical Statistics: Probability and
  Mathematical Statistics. John Wiley \& Sons, Inc., New York

\bibitem[Segers(2019)]{seg2019}
Segers J. 2019.
One- versus multi-component regular variation and extremes of {M}arkov trees.
Available from \texttt{https://arxiv.org/abs/1902.02226}.

\bibitem[Simpson et~al.(2018)Simpson, Wadsworth \& Tawn]{sim2018}
Simpson E, Wadsworth J, Tawn J. 2018.
Determining the dependence structure of multivariate extremes.
Available from \texttt{https://arxiv.org/abs/1809.01606}.

\bibitem[Smith et~al.(1997)Smith, Tawn \& Coles]{smi1997}
Smith R, Tawn J, Coles S. 1997.
Markov chain models for threshold exceedances.
\textit{Biometrika} 84:249--268

\bibitem[Smith(1992)]{smi1992}
Smith RL. 1992.
The extremal index for a {M}arkov chain.
\textit{Journal of Applied Probability} 29:37–45

\bibitem[Spirtes et~al.(2000)Spirtes, Glymour \& Scheines]{spi2000}
Spirtes P, Glymour C, Scheines R. 2000.
Causation, prediction, and search.
MIT Press, Cambridge, MA, 2nd ed.

\bibitem[Strokorb(2020)]{stro2020}
Strokorb K. 2020.
Extremal independence old and new.
Available from \texttt{https://arxiv.org/abs/2002.07808}.

\bibitem[Strokorb \& Schlather(2015)]{strokorb15}
Strokorb K, Schlather M. 2015.
An exceptional max-stable process fully parameterized by its extremal
  coefficients.
\textit{Bernoulli} 21:276--302

\bibitem[Tawn(1988)]{taw1988a}
Tawn JA. 1988.
Bivariate extreme value theory: Models and estimation.
\textit{Biometrika} 75:397--415

\bibitem[Thibaud et~al.(2016)Thibaud, Aalto, Cooley, Davison \&
  Heikkinen]{thi2015}
Thibaud E, Aalto J, Cooley DS, Davison AC, Heikkinen J. 2016.
{Bayesian inference for the {B}rown--{R}esnick process, with an application to
  extreme low temperatures}.
\textit{Ann. Appl. Stat.} 10:2303--2324

\bibitem[Varin et~al.(2011)Varin, Reid \& Firth]{var2011}
Varin C, Reid N, Firth D. 2011.
An overview of composite likelihood methods.
\textit{Statistica Sinica} 21:5--42

\bibitem[Wackernagel(2013)]{wac2013}
Wackernagel H. 2013.
Multivariate geostatistics.
Springer, New York.
An introduction with applications

\bibitem[Wadsworth \& Tawn(2012)]{wadsworth2012dependence}
Wadsworth JL, Tawn JA. 2012.
Dependence modelling for spatial extremes.
\textit{Biometrika} :asr080

\bibitem[Wadsworth \& Tawn(2014)]{wad2013}
Wadsworth JL, Tawn JA. 2014.
Efficient inference for spatial extreme value processes associated to
  log-{G}aussian random functions.
\textit{Biometrika} 101:1--15

\bibitem[Wadsworth et~al.(2017)Wadsworth, Tawn, Davison \&
  Elton]{Wadsworthetal2017}
Wadsworth JL, Tawn JA, Davison AC, Elton DM. 2017.
Modelling across extremal dependence classes.
\textit{Journal of the Royal Statistical Society: Series B (Statistical
  Methodology)} 79:149--175

\bibitem[Wainwright \& Jordan(2008)]{wainwright2008graphical}
Wainwright MJ, Jordan MI. 2008.
Graphical models, exponential families, and variational inference.
\textit{Foundations and Trends in Machine Learning} 1:1--305

\bibitem[Wan et~al.(2020)Wan, Wang, Davis \& Resnick]{wan2020}
Wan P, Wang T, Davis RA, Resnick SI. 2020.
Are extreme value estimation methods useful for network data?
\textit{Extremes} 23:171--195

\bibitem[Westra \& Sisson(2011)]{wes2011}
Westra S, Sisson SA. 2011.
Detection of non-stationarity in precipitation extremes using a max-stable
  process model.
\textit{Journal of Hydrology} 406:119 -- 128

\bibitem[Yu et~al.(2017)Yu, Uy \& Dauwels]{yu2017}
Yu H, Uy WIT, Dauwels J. 2017.
Modeling spatial extremes via ensemble-of-trees of pairwise copulas.
\textit{IEEE Transactions on Signal Processing} 65:571--586

\bibitem[Yuen \& Stoev(2014)]{yue2014}
Yuen R, Stoev S. 2014.
C{RPS} {M}-estimation for max-stable models.
\textit{Extremes} 17:387--410

\bibitem[Zhou(2010)]{zho2010}
Zhou C. 2010.
Dependence structure of risk factors and diversification effects.
\textit{Insur. Math. Econ.} 46:531--540

\bibitem[Zou et~al.(2019)Zou, Volgushev \& B{\"u}cher]{zou2019}
Zou N, Volgushev S, B{\"u}cher A. 2019.
Multiple block sizes and overlapping blocks for multivariate time series
  extremes.
Available from \texttt{https://arxiv.org/abs/1907.09477}.

\bibitem[Zscheischler \& Seneviratne(2017)]{zsc2017}
Zscheischler J, Seneviratne SI. 2017.
Dependence of drivers affects risks associated with compound events.
\textit{Science Advances} 3

\end{thebibliography}

\end{document}